\documentclass[11pt,a4paper]{article}
\pdfoutput=1
\usepackage[utf8x]{inputenc} 
\usepackage{jheppub}
\usepackage{amsfonts}
\usepackage{amsmath}
\usepackage{slashed}
\usepackage{lscape}
\usepackage{amssymb}
\usepackage{graphicx}
\usepackage{epic}
\usepackage{amsmath,amsfonts,amssymb,amsthm, bm}
\usepackage{multirow}
\usepackage{eepic}
\usepackage{epsfig}
\usepackage{latexsym}
\usepackage{color}
\usepackage{float}
\usepackage{multirow}
\usepackage{hyperref}
\usepackage{enumitem}
\hypersetup{colorlinks=true,citecolor=blue,linkcolor=magenta,urlcolor=blue}
\usepackage[caption=false]{subfig}
\usepackage{natbib}
\usepackage{relsize}
\usepackage{breqn}
\usepackage{shorthand}
\usepackage{subfig}
\usepackage[]{appendix}
\usepackage[normalem]{ulem}
\usepackage{soul}

\definecolor{darkred}{cmyk}{0,1,1,0.4}

\usepackage{color}
 \usepackage[normalem]{ulem}
\long\def\/*#1*/{}
\usepackage{color}
 \usepackage[normalem]{ulem}
\definecolor{darkgreen}{cmyk}{1,0,1,0.4}
\definecolor{darkred}{cmyk}{0,1,1,0.4}

\title{Retrieving Inverse Seesaw parameter space for Dirac Phase Leptogenesis}

\author[a]{Ananya Mukherjee,}
\emailAdd{ananyatezpur@gmail.com}

\author[b]{Nimmala Narendra,}
\emailAdd{nimmalanarendra@gmail.com}

\affiliation[a]{Theory Division, Saha Institute of Nuclear Physics, 1/AF Bidhannagar, Kolkata 700 064, India}
\affiliation[b]{Theoretical Physics Division, Physical Research Laboratory, Ahmedabad-380009, Gujarat, India}

\abstract{This work addresses the viability of \textit {Dirac phase leptogenesis}, in a scenario where the light Majorana neutrinos acquire masses by the inverse seesaw (ISS) mechanism. We show that, a successful leptogenesis in the ISS, driven (only) by the Dirac CP phase can be achieved with the involvement of an unorthodox form of the rotational matrix $R = e^{i{\bf A}} \,\,\,(e^{{\bf A}})$ in the Casas-Ibarra parametrisation. This particular structure of $R$ turns out to be an artefact in explaining the observed baryon asymmetry of the Universe in a pure ISS scenario. We detail here the confined regions of the $R$ matrix parameter space, essential for a successful leptogenesis. The $R$-matrix parameter space assists in rescuing the ISS parameter space needed for successful leptogenesis. This finding is otherwise unprecedented in the ISS set up. Making use of the resulted $R$ matrix parameter space we have calculated the branching ratio for the LFV decay $\mu \rightarrow e\gamma$. This accounts for an indirect probe of the $R$-matrix parameter space. The branching ratio obtained from the leptogenesis parameter space  surpasses the existing bound on the branching ratio that resulted in a scenario of combined effect of linear and inverse seesaw.  We also report here that, for $R = e^{i{\bf A}}$ choice leptogenesis demands the Dirac CP phase ($\delta$) to oscillate around $\pi/2$, although for the later choice the constraint on $\delta$ is much relaxed.} 

\begin{document}

\maketitle
\def\lapp{\mathrel{\rlap{\raise.5ex\hbox{$<$}}
                    {\lower.5ex\hbox{$\sim$}}}}
\def\gapp{\mathrel{\rlap{\raise.5ex\hbox{$>$}}
                    {\lower.5ex\hbox{$\sim$}}}}    

\section{Introduction}
Violation of the charge-parity~(CP) symmetry in the lepton sector has received attention in many aspects of low and high energy phenomena associated with non-zero neutrino mass and it's connection with matter-antimatter asymmetry. The aesthetic connection between low and high energy CP violation in the context of neutrino oscillation and leptogenesis ~\cite{Berger:1999bg,Pascoli:2006ci,Joshipura:1999is,Falcone:2000ib,Joshipura:2001ui,Rodejohann:2002hx,Davidson:2002em,Pascoli:2003uh,Frampton:2002qc,Branco:2002kt,Ellis:2002xg,Rebelo:2002wj,Endoh:2002wm,Endoh:2000hc}, however can be a manifest of the seesaw mechanisms offering neutrino mass. There has been plenty of theoretical activities in this direction, which emphasizes the effect of low energy CP violation on the CP violation that takes place at high energy (see Refs. ~\cite{Molinaro:2008rg,Moffat:2018smo}), realized by {\it leptogenesis}. In
this article we demonstrate such interplay considering the neutrino mass generation mechanism to be the inverse seesaw~(ISS). In this regard readers may look into the Ref.\,\cite{Dolan:2018qpy} where Dirac CP phase was assumed to be the only source coming from the low energy sector in order to engender a CP violation at high energy which is a potential idea of explaining the observed matter-antimatter asymmetry. Recently this difference has been reported to be $\eta_B = (6.04 \,-\, 6.2) \times 10^{-10}$  by the PLANCK collaboration \cite{Planck:2018vyg}.  Some time ago authors in \cite{PhysRevD.99.123508} proposed another way leading to successful {\it Dirac phase leptogenesis} with the consideration of a minimal flavor violation hypothesis (MLFV). Notwithstanding, we provide here an appealing alternate path leading to a successful {\it Dirac phase leptogenesis} in the pure ISS mechanism\footnote{We use the term ``pure" to highlight here the fact that, no other seesaw model has been considered in addition to the ISS.}. 

The ISS mechanism having a salient feature of offering the tiny neutrino mass at the cost of having a TeV scale heavy sterile states, makes a way to get itself verified in the collider experiments. Not only that, from theoretical perspective the ISS mechanism renders a large Yukawa coupling for a smaller mass window of the heavy sterile neutrinos \cite{Dias:2012xp}. Due to the TeV nature of the heavy RHNs present in the ISS, we are allowed to investigate the flavor effects in leptogenesis \cite{Blanchet:2006be} which may potentially lead to the dynamical generation of baryon asymmetry. Due to the large Yukawa couplings in ISS, it is difficult to have a successful leptogenesis given the source of CP asymmetry in the low energy is sourced by the Dirac CP phase only\footnote{As most of the asymmetries are seen to be erased by the huge washout (for detail see Ref.\,\cite{Dolan:2018qpy}).}. In this work we endeavour to provide a solution to the above issue leading to a successful {\it Dirac phase leptogenesis}. This is realized by considering an unorthodox choice of rotational matrix $R$ present in the Casas-Ibarra parametrization\footnote{Implication of this variety of the rotational matrix can be found in Refs.\,\cite{Pascoli:2003rq,Petcov:2005yh,Petcov:2006pc,Konar:2020vuu} and one of our companion work \cite{Mukherjee:2021hed}.}. It happens that, this particular class of the rotational matrix ($R$) controls over the lepton asymmetry production and the washout as well, both of which are guided by the Yukawa coupling which is constructed by $R$. 

The neutrino Yukawa coupling matrix, $Y^{\nu}$, together with the Majorana right handed neutrino~(RHN) mass matrix, $M_R$, and the charged lepton Yukawa coupling matrix, $Y^{\ell}$, constitutes one of the key ingredients both in the see-saw mechanism and in leptogenesis. In presence of a diagonal $Y^{\ell}$ and $M_R$ the $Y^{\nu}$ is found to be the only source of CP-violation in the lepton sector. For the simplistic scenario of type I seesaw, the neutrino Yukawa coupling matrix, $Y^{\nu}$ gets connected with the leptonic mixing matrix $U$~(also familiar as the PMNS matrix) through the relation $Y^{\nu}\, = \,(i/v)\sqrt{M_R} R \sqrt{m_n}U^\dagger$. This parametrization provided by Casas and Ibarra \cite{Casas:2001sr} hints the possible connection between the low and high energy CP violation. A detailed discussion on this has been kept in Section \ref{sec:cp}.

The primary goal of this work is to investigate whether leptogenesis in the ISS model can be solely guided by the Dirac CP phase or not. The reason behind this motivation is as follow. The feature of the ISS mechanism in the context of having a TeV scale heavy Majorana neutrino is significant in view of their production at the colliders (see \cite{Arun:2021yhm} and the references there in). On the other hand probing the Dirac CP phase in the current and upcoming neutrino oscillation experiments set enough motivation to further examine such issues. Motivated by this, we here bring forward a way to make the Dirac CP phase responsible for leptogenesis in the ISS model, assuming there to be no Majorana CP violation at all. To execute this idea we have considered two slightly different conventions for $R$ as $R = e^{i {\bf A}}$ and $R = e^{{\bf A}}$ rather the usual choice\footnote{By ``usual", we mean the widely used convention for the rotational matrix which in principle is comprised of three complex parameters. For simplicity which can be taken as real though (see e.g., \cite{Casas:2001sr}).}. More details on these choices of $R$ is provided in the following section. While executing the above goal we mention here the issues that one can encounter due to the present form of the rotational matrix $R$. These are respectively, 1) taking care of the relation among the different mass scales involved in the ISS model and the elements of the $R$ matrix, 2) determining the appropriate range of all these parameters which can lead to an adequate amount of lepton asymmetry that we are looking after and finally, 3) the validation of this particular $R$ matrix element in low energy experiments. The last point is taken care of while determining the branching ratio of LFV processes for instance $\mu \rightarrow e \gamma$. We find that this branching ratio obtained from the leptogenesis parameter space surpasses the existing bound on the branching ratio that resulted in a scenario of combined effect of linear and inverse seesaw \cite{Dolan:2018qpy}.  Analysis on the role of this $R$ matrix can be found in the context of leptogenesis~\cite{Pascoli:2006ci} and lepton flavor violation~\cite{Pascoli:2003rq} in a scenario where neutrinos get masses by type-I seesaw mechanism.  In a companion work \cite{Mukherjee:2021hed} we have assessed this idea for type I seesaw mechanism and commented on the required range of Dirac CP phase for flavored leptogenesis which is assumed to take place at $10^8$ GeV. In addition to the case of having a Dirac CP violation we also investigate the case where an absence of low energy CP violation\footnote{The condition when both Dirac and Majorana CP violations are turned off.} is seen to give rise to the observed baryon asymmetry of the Universe, which is discussed in the Appendix. However, such impact of low energy CP violation or conservation does not get reflected in the phenomenology related to the concerned LFV processes.

We organize this article as follows. In section \ref{sec:CI} we brief on the ISS mechanism and the necessary extraction of Yukawa coupling through the CI formalism. Section \ref{sec:lepto} provides the explanation for baryogenesis through leptogenesis in ISS scenario. Constraints on the ISS parameter space from $\mu \rightarrow e \gamma$  have been provided in section \ref{sec:lfv}. The overall phenomenology we discuss in section \ref{sec:analysis}. Finally we conclude in section \ref{sec:conclusion}. The appendix contains the overall phenomenology related to the restrictions on the involved parameters, in absence of the Dirac CP violation. 
  
\section{Inverse seesaw Yukawa coupling}\label{sec:CI}
The extended type-1 seesaw mechanism, whose new physics can be essentially manifest at the TeV scale is familiar as the ISS mechanism proposed in Ref.~\cite{PhysRevLett.56.561,PhysRevD.34.1642}. The key feature of this seesaw realization comes through the violation of lepton number at a low energy scale and, as a bonus it hints a new physics at the TeV scale.
With the small lepton number violating mass scale ($\mu$ of the order of {\it keV}), the ISS mechanism offers a sub-eV ordered neutrino mass at the cost of trading a set of SM gauge singlet (denoting as $S_L$) fermions along with the traditional three copies of TeV scale RHN ($N_{R}$). The following Lagrangian describes such mass generation mechanism of light neutrinos under the ISS mechanism. 
\begin{equation}\label{ISS_Lag}
-\mathcal{L} \supset \lambda_{\nu}^{ \ell i} \,\overline{L_\ell} \,\widetilde{H} \,N_{R_i} + M_{R} \,\overline{(N_{R_i})^{c}} \,S_{L_i}^c + \frac{1}{2} \mu \overline{S_{L_i}} (S_{L_i})^c + h.c.
\end{equation}
with, $\widetilde{H} = i \sigma_2 H^*$. The $\ell,i$ being respectively the flavor and generation indices of leptons, RHNs and newly added SM gauge singlet fermions. The complete mass matrix constructed from the basis $( \nu_L^c, N_R, S_L^c)$ with the help of Eq.~\ref{ISS_Lag} can be written in the following texture:
\begin{equation}
M_\nu = \begin{pmatrix}\label{ISS_matrix}
0 && m_D && 0 \\
m_D^T && 0 && M_R \\
0 && M_R^T && \mu \\
\end{pmatrix},
\end{equation}  
where, $m_D,\, M_R\,\, \text{and}\,\, \mu$ are $3\times3$ mass matrices. The light neutrino mass can be found by going through a block diagonalization of the above matrix, which leads to the following
\begin{equation}\label{eq:mnu}
m_\nu = m_D (M_R^T)^{-1}\mu M_R^{-1}m_D^T.
\end{equation}
The $6\times6$ neutrino mass matrix formed with the bases $(N_R, S)$ can be put into the following form,
\begin{equation}
M_\nu^{6\times6} = \begin{pmatrix}\label{6by6}
0 && M_R \\
M_R^T && \mu  \\
\end{pmatrix},
\end{equation}
Diagonalizing the above one can write the resulting pseudo-Dirac mass states having the following form~\cite{Dolan:2018qpy},
\begin{equation}\label{eq:splitting}
M_N =  \frac{1}{2} \left( \mu \pm \sqrt{\mu^2 + 4M_R^2}\right),
\end{equation} 
with $\mu$ as the same lepton number violating scale which essentially acts as the source of tiny non-degeneracy among the final pseudo-Dirac pairs in the ISS model. 

The extraction of the ISS Yukawa coupling through the CI parametrization is obtained as \cite{Dolan:2018qpy},
\begin{equation}
\lambda_{\nu} = \frac{1}{v} \,U \,m_{n}^{1/2} \,R \, \mu^{-1/2}\,M_{N}^T\,,
\end{equation}
where $\,m_n,\,M_{N}$ and $\mu $ are the $3\times3$ mass matrices having definitions $m_n \equiv \text{diag}(m_{1},m_{2},m_{3})$ and $M_{N} \equiv \text{diag}(M_{1},M_{2},M_{3})$. Without loss of generality, we assume $\mu$ also to be diagonal. Here $v$ denotes the SM Higgs vacuum expectation value (VEV). In general $R$ is a complex orthogonal matrix, $R R^{T}={\bf \mathbb{I}}$. This orthogonality condition permits us to have two additional different choices for $R$: i) $R=  O \,e^{i {\bf A}}$ and ii) $R= O \,e^{{\bf A}}$, where ${\bf A}$ is a skew symmetric matrix. For simplicity we choose $O$ to be an identity matrix\footnote{By definition  $O$, can be a complex orthogonal matrix, which however does not change the related phenomenology of this analysis much.} 

 The diagonalization of the light neutrino mass matrix is followed by writing: $U^{\dagger} \,m_{\nu} \,U^{*}=m_\nu ^{\text{diag}}$, where $U$ being the lepton mixing matrix $U\simeq U_{\text{PMNS}}$, having the following form,
\begin{equation}
U=\left(\begin{array}{ccc}
c_{12}c_{13}& s_{12}c_{13}& s_{13}e^{-i\delta}\\
-s_{12}c_{23}-c_{12}s_{23}s_{13}e^{i\delta}& c_{12}c_{23}-s_{12}s_{23}s_{13}e^{i\delta} & s_{23}c_{13} \\
s_{12}s_{23}-c_{12}c_{23}s_{13}e^{i\delta} & -c_{12}s_{23}-s_{12}c_{23}s_{13}e^{i\delta}& c_{23}c_{13}
\end{array}\right) U_{\text{M}}.
\label{PMNS}
\end{equation}
We define $c_{ij} = \cos{\theta_{ij}}, \; s_{ij} = \sin{\theta_{ij}}$ as the three mixing angles and $\delta$ as the Dirac CP phase. The diagonal matrix $U_{\text{M}}=\text{diag}(1, e^{i\alpha_{1}}, e^{i \alpha_{2}})$ contains the undetermined Majorana CP phases $\alpha_{1}, \, \alpha_{2}$. One can simply express the diagonal light neutrino mass eigen values in terms of the solar and atmospheric mass squared splittings\footnote{$m^{\text{diag}}_{\nu}  = \text{diag}(m_1, \sqrt{m^2_1+\Delta m_{21}^2},\sqrt{m_1^2+\Delta m_{31}^2})$ for normal hierarchy (NH) and $m^{\text{diag}}_{\nu} = \text{diag}(\sqrt{m_3^2+\Delta m_{23}^2-\Delta m_{21}^2}$, $\sqrt{m_3^2+\Delta m_{23}^2}, m_3)$ for inverted hierarchy (IH).}. A recent status of the neutrino oscillation parameters can be found in Ref.~\cite{Esteban:2018azc}. For the numerical simulation we use the best fit central values of the oscillation parameters as below:
\begin{gather}\label{eq:nudata}
\sin^2\theta_{23} = 0.582,\, \sin^2\theta_{13} = 0.0224, \,\sin^2\theta_{12} = 0.310\\ \nonumber
\Delta m^2_{21} =  7.39\,\times\,10^{-5} \text{eV}^2, \Delta m^2_{31}= 2.525\, \times\,10^{-3}\text{eV}^2.
\end{gather}
The above values of oscillation parameters correspond to the normal mass ordering of the light neutrino mass eigenvalues. Along with the global fit oscillation parameters we use the lightest neutrino mass to be $10^{-3}$~eV through out this analysis\footnote{Mentioning that a different value of the lightest neutrino mass from the range $ 10^{-5} \leq m_{\rm lightest}/ \rm{eV} \leq 0.05$ does not bring significant changes in the success of {\it Dirac phase leptogenesis} in ISS scenario.}. It is also to note here that we restrict the entire analysis considering the neutrino masses to be normally ordered.

We learn that, the difference between the neutrino and anti-neutrino oscillation probability imply a CP violation in the neutrino oscillation phenomenon. And also this difference is sensitive to the Dirac CP phase. Several baseline and reactor experiments are in order for the measurement of the $\delta$. For a few recent investigations on the $\delta$ value one can look into Refs.~\cite{Abe:2019vii,Acero:2019ksn}. Very recently {\bf T2K} has reported a maximal CP violation which sets this value around ($- \pi/2$) for both mass orderings. Although the analysis done by {\bf NO}$\nu${\bf A} excludes most values near $\delta 
\,=\, \pi/2$ for the inverted mass ordering by more than $3\sigma$. Keeping this experimental findings in mind, we further investigate the $\delta$ range from the requirement of having a $\delta$ driven leptogenesis.
\subsection{CP phases in the ISS Yukawa coupling}\label{sec:cp}
The idea of leptogenesis induced by the low energy phases is envisaged through their appearance in the Yukawa coupling which governs the concerned decay for leptogenesis. As we will see below there may remain more than one such phase which can ensure the complex nature of the Yukawa coupling. This discussion thus can lead to the following couple of sentences.
The source of the essential CP-violation in leptogenesis can, in principle, be more than one which are respectively, the rotational matrix $R$, the PMNS matrix $U$, or $R$ and $U$ together. However, identification of the CP-violation due to the PMNS phases and that due to the $R$ matrix at high energies, e.g. in leptogenesis, requires some explanation. If the $R$ matrix elements are purely real or imaginary \cite{Pascoli:2006ci}, along with the assumption that, neither Dirac nor Majorana CP violation is taking place, it results no CP violation at high energy, which means no leptogenesis at all. In an alternate scenario, such as in presence of Dirac and/or Majorana CP violation leptogenesis can be possible even if one goes with the choices of a purely real or imaginary $R$. This scenario can be considered as leptogenesis driven by the low energy CP phases, present in the PMNS matrix. In the following, we will be familiar with all the possible sources producing a complex Yukawa coupling matrix and consequently one or more CP violating sources. 

\begin{center}\textbf{Case-I: $R=e^{i{\bf A}}$}\end{center}
With the above choice of $R$ matrix the Yukawa coupling matrix can be written as, 
 \begin{equation}\label{eq:CI_1}
\lambda_{\nu} = \frac{1}{v} \,U \,m_{n}^{1/2} \,e^{i{\bf A}} \,\mu^{-1/2}\,M_{N}^{T}.
\end{equation}
In case of this complex parameterisation the orthogonal matrix $R=e^{i {\bf A}}$ can be expanded as~\cite{Pascoli:2003rq,Petcov:2006pc,Mukherjee:2021hed}:
\begin{equation}
e^{i {\bf A}}=1-\frac{\cosh r-1}{r^{2}} {\bf A}^{2}+i\frac{\sinh r}{r} {\bf A},
\label{eiA_expr}
\end{equation}
where, ${\bf A}$ is a real skew symmetric matrix having satisfied the feature of the $R$ matrix to be orthogonal ($R R^T = \mathbb{I}$). 
\begin{equation}
\begin{pmatrix}\label{skew}
0 && a && b \\
-a && 0 && c \\
-b && -c && 0 \\
\end{pmatrix},
\end{equation}           
with, $r=\sqrt{a^{2}+b^{2}+c^{2}}$. Owing to the presence of the imaginary quotient $i$, and the assumption $O \approx I$ (rather being comprised with three complex angles) we often call the case with $R = e^{i {\bf A}}$ as the complex case and the later (with $R=e^{{\bf A}}$) as the real case. In here with this choice of $R$, we have two potential sources of CP violation for instance the element of $R$-matrix itself and $\delta$. Later we will see how both of them play pivotal roles in finally achieving the observed baryon to photon ratio.  

 \begin{center}\textbf{Case-II: $R=e^{{\bf A}}$ }\end{center}
 The Yukawa matrix has the following structure with the choice $R=e^{{\bf A}}$
 \begin{equation}\label{eq:CI_2}
\lambda_{\nu} = \frac{1}{v} \,U \,m_{n}^{1/2} \,e^{{\bf A}} \,\mu^{-1/2}\,M_{N}^{T}.
\end{equation}
In case of this parameterisation the orthogonal matrix $R=e^{{\bf A}}$ can be expanded in the following form \cite{Pascoli:2006ci,Mukherjee:2021hed}:
\begin{equation}
e^{{\bf A}}=1+\frac{1-\cos r}{r^{2}} {\bf A}^{2} + \frac{\sin r}{r} {\bf A},
\label{eA}
\end{equation}
where, the matrix {\bf A} and $r$ have the similar definitions as in the former case described above. 
 In order to proceed with a minimal number of free parameters, we assume here the elements of the matrix ${\bf A}$ to obey $a=b=c= \, \kappa$, and hence we write $r=\sqrt{3}\, \kappa$\footnote{This kind of equality among these three elements can be found in a very recent work \cite{Konar:2020vuu,Mukherjee:2021hed}.}. Henceforth, in this analysis we mention the rotational matrix parameter space by referring to the parameter {\bf $\kappa$} only.  From Eq.~\ref{eq:CI_2} one can notice that with this choice for $R$-matrix there exist only one potential source of CP violation that is the Dirac CP phase. Thus for a {\it Dirac Phase leptogenesis} to occur with $R=e^{{\bf A}}$ one must have low energy CP violation which is assumed to bring out via $\delta$ which appears in the PMNS matrix.

\subsection{Deviation from unitarity of the $U_{\text{PMNS}}$} 
Owing to the assumed unitarity of the PMNS matrix, there has been theory describing the ``no-connection" between the low and high energy CP violation \cite{Rodejohann:2010zz}, provided the choice of RHN mass falls at or above the canonical type-I seesaw scale\footnote{Because, leptogenesis is assumed to take place at the scale of the right handed neutrino mass. However, this finding no longer holds good for high scale leptogenesis induced by the low energy CP violation \cite{Moffat:2018smo}. And also, leptogenesis below a particular mass scale (at least $10^{11}$~GeV) let this no-connection theorem go off. As, this temperature regime dictates the lepton flavors to remain into equilibrium. And also in a low scale seesaw model like ISS, the presence of the TeV scaled RHN makes the non-unitarity parameters larger in comparison to what one estimates for type-I seesaw having RHN mass of the order of $10^{12}$ GeV.}. On the other hand, this connection can be attributed to the presence of a large magnitude of Non-Unitarity (NU) parameters, which is a manifest of any low scale seesaw model. Non-unitarity appears whenever additional heavy particles mix with the light neutrinos.

As stated above, the diagonalization of $m_\nu$ by the PMNS matrix does not diagonalize the $M_R$ and $\mu$. There remains off-diagonal terms arising from the mixing among the light neutrinos even after the diagonalization of $m_\nu$ due to their mixing with the heavy neutrinos. To rephrase it, in a basis where the charged-lepton mass matrix is diagonal, $U$ is only a part of the full mixing matrix realizing neutrino oscillations. As a result, the usual PMNS matrix (Eq.~\ref{PMNS}) in principle, gets replaced by a non-unitary mixing which can be denoted as $N$, and hence can be realized by writing, $N = (1\,-\,\eta)\,U = (1\,-\, \Theta \Theta^\dagger /2) \,U$ \cite{Blennow:2016jkn}. Here, $\eta$ stands for the NU parameter measuring the deviation of the PMNS matrix from Unitarity. Order of the NU in the ISS can be obtained by the ratio of $(m_D M_R^{-1})$, which turns out to be large for a TeV $M_R$ and a few GeV $m_D$. Henceforth, in this analysis we will use the NU mixing matrix ($N$) instead of $U$ as given by Eq.~\ref{PMNS} for the extraction of Yukawa coupling and wherever needed. In the following we brief the discussion related to the NU mixing obtained in the ISS model, which are going to be further utilised in the calculation of lepton asymmetry and branching of LFV decays.     
The full mass matrix of ISS ($M_\nu^{9 \times 9}$) in general, is of dimension $9 \times 9$ which can be derived from the basis ($\nu_i, M_{Ri}, \,\text{and}\,\,S_{Li}$). One can express the diagonalising matrix of $M_\nu^{9 \times 9}$ as, 
\begin{equation}
V = \left( \begin{array}{cc}V_{3 \times 3} &V_{3 \times 6} \\
V_{6 \times 3} & V_{6 \times 6}\end{array}\right)
\end{equation}
with, $M_\nu^{\rm diag} = V^T M_{\nu} V$. In here, the upper-left sub-block $V_{3 \times 3}$ represents the full non-Unitary  mixing matrix. 
In order to diagonalise the $9\times9$ neutrino mass matrix, in presence of the NU, one can consider the following structure of the mixing matrix $V$,
\begin{equation}
V = \left( \begin{array}{cc}  (1_{3\times3} + \zeta^* \zeta^T)^{-1/2} & \zeta^*(1_{6\times6} + \zeta^T\zeta^*)^{-1/2}\\
-\zeta^T(1_{3\times3} + \zeta^*\zeta^T)^{-1/2} & (1_{6\times6} + \zeta^T \zeta^*)^{-1/2}\end{array}\right)  \left(\begin{array}{cc}U & 0\\0& V^\prime \end{array}\right)
\end{equation}
With the assumption of a minimal flavor violation implying $M_R$ and $\mu$ to be diagonal, $V^\prime$ can be evaluated as,
\begin{equation}
V^{\prime} = \frac{\sqrt{2}}{2} \left(\begin{array}{cc} 1_{3 \times3} & -i 1_{3 \times3} \\ 
1_{3 \times 3}& i 1_{3 \times 3}\end{array} \right)+ \mathcal{O} (\mu M_R^{-1})
\end{equation}
At the same time, $\zeta$ follows a definition like, $\zeta = (0 _{3\times3}, m_D M_R^{-1})$. It is now clear that, each component of $V$ can be expressed with $\zeta$ and $m_D M_R^{-1}$ as written below,
\begin{equation}
V_{3\times 3} = \Big(1_{3\times3} + \zeta^* \zeta^T\Big)^{-1/2} U \simeq \Big(1_{3\times3} - \frac{1}{2}\Theta^* \Theta^T \Big) U = (1_{3\times3} - \eta) U
\end{equation}
Where, $\Theta = m_D M_R^{-1}$ as mentioned earlier, $\eta$ measures the deviation from Unitarity  having it's form, $\eta \equiv  1_{3\times3}-(1_{3\times3} + \zeta^* \zeta^T)^{-1/2}$. Thus it is evident that, $\Theta = m_D M_R^{-1} << 1_{3\times3}$ can be achieved in ISS. 

Now it is to understand the role of $\kappa$ and $\mu$ while taking care of the magnitude of the NU (here we call it by $\Theta$). As set by the traditional ISS mass scales for $m_D,\, M_R, \text{and} \,\, \mu$ the NU scale $\Theta$ should fall $\leq 10^{-2}$ . From Eqs. \ref{eq:CI_1} and \ref{eq:CI_2} one can obtain $m_D$ as a function of $\kappa$ and $\mu$. For both the choices of $R$-matrix we see an exponential rise in $\kappa$ which may encounter large values of $m_D$. Especially for $R\,=\, e^{ i{\bf A}}$ the enhancement in the order of $m_D$ is huge for large $\kappa$ values. At the same time, $m_D$ is inversely proportional to $\sqrt{\mu}$. Thus one should be careful enough while playing with the $\mu$ scale, as an increase or decrease from the traditional ISS scale which is $\mathcal{O}(keV)$, can disturb the NU scale. This leads us to further constrain the $\kappa$ parameter space as we will see in the Sec. \ref{sec:lfv}.

\section{Leptogenesis in inverse seesaw}\label{sec:lepto}
In the ISS model the decay of the pseudo-Dirac neutral states trigger the generation of lepton asymmetry we are looking for. These pseudo-Dirac neutral states (denoted by $N_k$) undergo CP violating decay into the SM lepton ($L_{\ell}$) and the Higgs doublet ($H$) as, 
    \begin{equation}
    \epsilon_{N_{k}}^\ell = -\sum \frac{\Gamma(N_{k} \rightarrow L_{\ell}+H^{+} , \nu_l+ H^0)-\Gamma(N_{k} \rightarrow L_{\ell}+H^-, \nu_\ell^c +H^{0^*})}
    {\Gamma(N_{k} \rightarrow L_{\ell}+H^{+} , \nu_\ell +H^0)+\Gamma(N_{k} \rightarrow L_{\ell}+H^{-}, \nu_\ell^c +H^{0^*})}
   \end{equation} 
As evident from above the CP-asymmetry is a measure of the difference in decay widths of $N_k$ to a process and its conjugate process. At the tree level, these two are the same giving rise to vanishing CP-asymmetry. Taking into account the one loop vertex and self energy diagrams, it is found that non-zero CP-asymmetry arises due to the interference between the tree level and the one loop diagrams. For the decaying pseudo-Dirac mass falling in the TeV regime, leptogenesis relies on a mechanism, in which self-energy effects on the leptonic CP asymmetry become dominant and get resonantly enhanced. This happens when the pseudo-Dirac states are nearly degenerate in their masses and the mass splitting is comparable to the pseudo-Dirac decay width \cite{Pilaftsis:2003gt,Pilaftsis:1997jf,Flanz:1996fb,Xing:2006ms}.

As we learn, leptogenesis at TeV scale recasts the idea of flavored leptogenesis (please see Refs.~\cite{Blanchet:2006be,Abada:2006ea,Nardi:2006fx}, for a detailed discussion). It happens that at such low temperature regime all the lepton flavors come in equilibrium with the standard model plasma and are distinguishable from each other. Different lepton
flavors can in general be identified by their Yukawa couplings $Y_\nu^\ell$ ($\ell = e,\mu~\text{and}~\tau$), ensuring that $\ell$-th lepton flavor becomes distinguishable when $Y_\nu^\ell$ related interactions enter into equilibrium.  And this is examined by comparing the $Y_\nu^\ell$ related
interaction rates with the Hubble expansion rate of the Universe, at a particular temperature we are interested in. The flavor effects in leptogenesis also have considerable impact on the investigation of the washout due to individual lepton flavors as the former depends on $Y_\nu^\ell$. 

 We find it instructive to mention that one must pursue change-of-basis exercise for the Yukawa couplings that are going to be used for the calculations of $\epsilon$ and $\eta_B$.  In concord with the fact that, the final pseudo-Dirac states only take participation in the leptogenesis decay, this is essential to be carried out. One can find the procedure adopted in \cite{Dolan:2018qpy} in order to choose the appropriate bases for the pseudo-Dirac pair mass ($M_N$) and the Yukawa couplings ($Y_\nu$).

As evident from Eq.~\ref{eq:splitting}, ISS model has the feature of rendering a natural way of keeping the pseudo-Dirac states nearly degenerate, as required by the condition of having resonant leptogenesis. However, an exactly degenerate pseudo-Dirac mass states will result into a vanishing lepton asymmetry as clear from the Eq.~\eqref{eq:asymmetry}. As mentioned earlier, flavor-dependent effects of leptogenesis are relevant at low enough temperatures (set by the RHN mass) such that at least one charged lepton Yukawa coupling is in thermal equilibrium. When this condition is met, flavor-dependent effects are not avoidable as the efficiency factors differ significantly for the distinguishable flavors. The relevant expression for lepton asymmetry parameter, which is obtained by taking the effects of all the lepton flavors is given by~\cite{Deppisch:2010fr,Bambhaniya:2016rbb,Dev:2017trv},
 \begin{equation}\label{eq:asymmetry}
    \epsilon_i^ \ell = \sum_{j \neq i} \frac{\text{Im}[Y_{\nu_{i\ell}}Y_{\nu_{j\ell}}^{*}(Y_{\nu}Y_{\nu}^\dagger)_{ij}]+\frac{M_i}{M_j}\text{Im}[Y_{\nu_{i\ell}} Y_{\nu_{j\ell}}^{*}(Y_{\nu}Y_{\nu}^\dagger)_{ji}]}{(Y_{\nu}Y_{\nu}^{\dagger})_{ii}(Y_{\nu}Y_{\nu}^{\dagger})_{jj}} f_{ij}^{mix}   
   \end{equation}
with the regulator given by,
\begin{equation*}
f_{ij}^{mix} = \frac{(M_{i}^2 - M_j^2)M_{i}\Gamma_{j}}{(M_{i}^2 - M_j^2)^2 + M_i^2 \Gamma_j^2}  
\end{equation*}
 with $\Gamma_i = \frac{M_i}{8\pi}(Y_\nu Y_\nu^\dagger)_{ii}$ as the tree level decay width of the pseudo-Dirac state. There is also a similar contribution ${\epsilon_i^\ell}^{osc}$ to the CP-asymmetry emerging from the oscillation between the pseudo-Dirac states ~\cite{Blanchet:2010kw,Kartavtsev:2015vto,Dev:2015wpa}. Its expression is given by Eq.~\eqref{eq:asymmetry} with the replacement $f_{ij}^{mix} \rightarrow f_{ij}^{osc}$, where 
\begin{equation*}
f_{ij}^{osc} = \frac{(M_{i}^2 - M_j^2)M_{i}\Gamma_{j}}{(M_{i}^2 - M_j^2)^2 + (M_i \Gamma_i +M_j \Gamma_j)^2 \frac{\text{det}[\text{Re}(Y_\nu Y_\nu^\dagger)]}{(Y_\nu Y_\nu^\dagger)_{ii}(Y_\nu Y_\nu^\dagger)_{ii}}}  
\end{equation*}

With the above prescription for lepton asymmetry one can write the analytically approximated solution for the baryon to photon ratio \cite{Pilaftsis:2003gt,Deppisch:2010fr} as,
 \begin{equation}\label{Eq:bau}
 \eta_B \simeq -3\times 10^{-2} \sum_{\ell,i}\frac{\epsilon_i ^\ell}{K_\ell^{\text{eff}}\text{min}\left[z_c,1.25 \, \text{Log}\,(25 K_\ell^{\text{eff}})\right]} ~,
\end{equation}  
where $z_c = \frac{M_i}{T_c}$ and $T_c \sim 149$ GeV, \cite{Bambhaniya:2016rbb} is the critical temperature, below which the sphalerons freeze out \cite{PhysRevD.49.6394,DOnofrio:2012phz}. Here,  $K_\ell^{\text{eff}} = \kappa_\ell \sum_{i} K_i B_{i\ell}$ , with $K_i = \Gamma_i /H$, the wash out factor and $\Gamma_i = \frac{M_i}{8\pi}(Y_{\nu} Y_{\nu}^\dagger)_{ii}$ as the tree level heavy-neutrino decay width. The Hubble rate of expansion at temperature $T\sim M_i$ can be expressed as ,
\begin{equation*}
H = 1.66 \sqrt{g^*}\frac{M_i^2}{M_{\text{Pl}}}\;\; \text{with}\;\;\; g^* \simeq 106.75 \;\;\;\text{and}\;\;\; M_{\text{Pl}} = 1.29 \times 10^{19} \,\text{GeV}.
\end{equation*}
 Here, $B_{i\ell}$'s are the branching ratios of the $N_i$ decay to leptons of $\ell^{th}$ flavor : $B_{i\ell} = \frac{|Y_{\nu_{i\ell}}|^2}{(Y_{\nu}Y_{\nu}^{\dagger})_{ii}}$.
Including the Real Intermediate State~(RIS) subtracted collision terms one can write the factor $\kappa$ as,
\begin{equation}
\kappa_\ell = 2 \sum_{i,j j \neq i} \frac{\text{Re}\left[(Y_{\nu})_{i\ell}(Y_{\nu})_{ j\ell}^* \left(Y_{\nu} Y_{\nu}^\dagger\right)_{ij}\right]+ \text{Im}\left[\left(\left(Y_{\nu}\right)_{ i\ell} (Y_{\nu})_{ j\ell}^*\right)^2\right]}{\text{Re}[(Y_{\nu}^\dagger Y_{\nu})_{\ell \ell} \{(Y_{\nu} Y_{\nu}^\dagger)_{ii} + \left(Y_{\nu} Y_{\nu}^\dagger\right)_{jj}\}]}\times\left(1-2i \frac{M_i-M_j}{\Gamma_i + \Gamma_j}\right)^{-1}.
\end{equation} 
The above expressions for lepton asymmetry, washout and efficiency factors evince that, all of them explicitly depend on the Yukawa coupling which follows the construction as given by Eq.~\ref{eq:CI_1} and  Eq.~\ref{eq:CI_2} respectively for the two choices of $R$ matrix. Before going to the calculation of lepton asymmetry, we derive the relations among the mass scale of the pseudo-Dirac pairs and the parameter which builds the $R$-matrix (denoted as $\kappa$). The detailed methodology of which we discuss in the following section.  
\section{Constraints on parameters from $|Y_\nu| \leq \sqrt{4 \pi}$}\label{sec:analysis}
As already mentioned, the introduction of an unorthodox structure of the rotational matrix in the CI formalism makes the leptogenesis scenario in the ISS framework phenomenologically viable. The construction of the Yukawa coupling that governs the decay of the pseudo-Dirac state and also the generation of neutrino mass is followed by the involvement of this special $R$ matrix forms. The exponential dependency on $\kappa$ of the Yukawa couplings makes a sharp rise in it's order of magnitude. For $R= e^{i A}$ this dependency is hyperbolic however, for $R= e^{A}$ it is sinusoidal. The free parameters involved in the extraction of Yukawa coupling from Eqs.~\ref{eq:CI_1} and \ref{eq:CI_2} thus can be noted as given below,
$$ \{ \kappa, M_N, \mu\}.$$
\begin{figure*}[h]
\begin{center}
\includegraphics[scale=0.28]{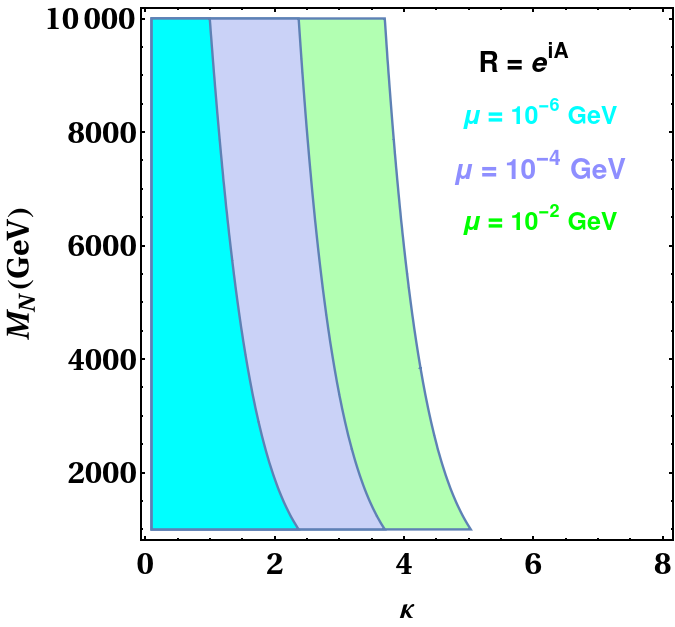}
\includegraphics[scale=0.28]{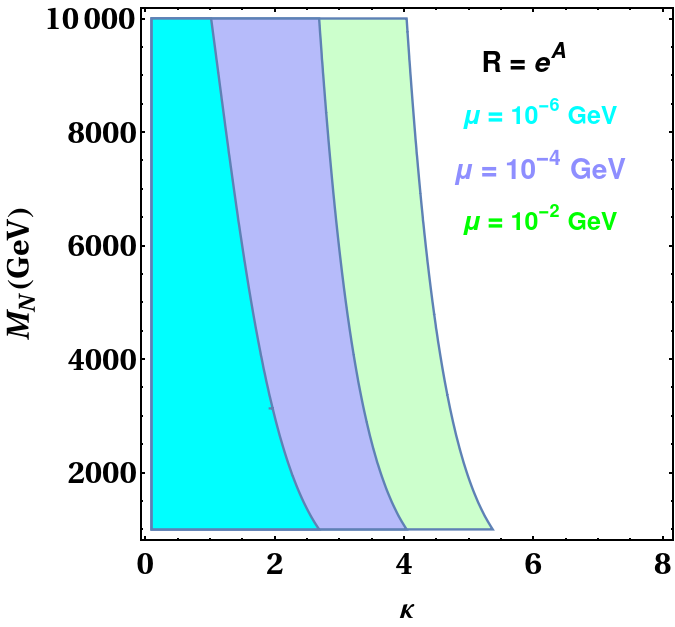}  
\caption{Shows the allowed parameter space for the matrix element $\kappa$ and the pseudo-Dirac mass $M_N$ in the ISS for $|Y_\nu| \leq 1$. Different color codes have been used to identify the allowed regions satisfying the former constraint corresponding to the chosen value of $\mu$.} 
\label{fig:kappa}
\end{center}
\end{figure*}
The value of $\kappa$ can not be arbitrary in the sense that, a large value of $\kappa$ can result into $|Y_\nu| > \sqrt{4\pi}$, breaking the perturbativity limit. Hence, to be in agreement with the criteria $|Y_\nu| \leq \sqrt{4\pi}$, one should be careful of selecting the relevant range for $\kappa$. For this, we evaluate the allowed ranges of $\kappa$ and the overall mass scale for the neutral pseudo-Dirac states satisfying the aforementioned criteria. While doing so we have considered different benchmark values of the LNV scale ($\mu$) and found that the allowed region of $\kappa$ is more sensitive to the former. The obtained ranges of $\kappa$ and $M_N$ are presented in the form of region plots as described by Fig.~\ref{fig:kappa}. It is worth noting that, different $\mu$ scales can significantly impact the choice of $\kappa$ values. Keeping that in mind, we have estimated the allowed regions for $\kappa$ considering $\mu = 10^{-6,\,-4,\,-2}$~GeV. For each choices of the $R$ matrix the range of $\kappa$ turns out to be a little different as evident from this figure. The above region plots correspond to the following relations (Eqs. \ref{eq:kappa_mr1} and \ref{eq:kappa_mr2}) between $\kappa$, $M_N$ and $\mu$ subject to each choice of the $R$ matrix. Making use of Eqs.~\ref{eq:CI_1} and \ref{eq:CI_2} along with the neutrino data from Eq.~\ref{eq:nudata} one can derive the following analytical expressions which relate the parameters $\mu$ and $M_N$ scale with $\kappa$.
\begin{equation} \label{eq:kappa_mr1}
 \text{When} \,\,R= e^{iA}: \,\, \text{Cosh}[\sqrt{3}\,\, \kappa] \leq \Big | \left(153.744 \sqrt{-\frac{\mu }{\text{$m_\nu $} M_N^2}}+\frac{272.598 \sqrt{\mu }}{\sqrt{\text{$m_\nu $}} M_N}\right)\,\, \text{GeV} -0.173136 \Big|\end{equation}

\begin{equation}\label{eq:kappa_mr2}
  \text{When}\,\, R= e^{A}:  \,\, \text{Cosh}[\sqrt{3}\,\, \kappa] \leq  \Big| \left(477.809 \sqrt{-\frac{\mu }{\text{$m_\nu $} M_N^2}}-\frac{275.863 \sqrt{\mu }}{\sqrt{\text{$m_\nu $}} M_N} \right)\,\, \text{GeV}+1.19297 \Big|
  \end{equation}
 It makes us convenient to write from Fig.~\ref{fig:kappa} that for larger $\mu$ values one can have wider allowed region for $\kappa$ as imposed by the condition $Y_\nu \leq \sqrt{4 \pi}$. 
As depicted by Fig.~\ref{fig:kappa}, these ranges of $\kappa$ and $M_N$ are going to be used for the calculation of the branching ratio for $ \mu \rightarrow e \gamma $ and baryon asymmetry as presented in the following sections. For making these plots we have used the neutrino data as provided in Eq.~\ref{eq:nudata}. 

\section{Results and discussion}
\subsection{Constraints from $\mu \rightarrow e  \gamma$}\label{sec:lfv}
The influence of these $R$ matrices (Eq.~\ref{eq:CI_1} and \ref{eq:CI_2}) in rising the order of magnitude of the branchings of various LFV processes, have been emphasized earlier in Refs.~\cite{Pascoli:2003rq} and very recently in \cite{Mukherjee:2021hed} for type-I seesaw scenario. In here, we have performed the analysis for the ISS framework with these two structures of the $R$-matrix. It is thus convenient to write that, the enhancement in the order of the branching ratio is the outcome of the exponential dependency of the light-heavy mixing ($V_{\alpha i}$)
on the parameter $\kappa$ that constructs the $R$-matrix. Here, we have focused only on the study of the particular BR for $ \mu \rightarrow e  \gamma $ decay process, which presently provides the strongest bound in comparison to other variants of LFV decay. In the ISS scenario, one can naturally obtain a large branching of these LFV decays in comparison to what one obtains in the type-I seesaw mechanism\footnote{As we have seen, the rates of the LVF processes in the canonical type-I seesaw model with massive right handed neutrinos are so strongly suppressed that these processes are not observable in practice, and one has {\it e.g.} BR$(\mu \rightarrow e  \gamma ) < 10^{-47}$ \cite{Cheng:1980tp,Aubert:2009ag} in a type-I seesaw scenario.}. This large BR (here in particular, BR$(\mu \rightarrow e \gamma)$) is in practice resulted from the large light-heavy mixing (denoted by $V_{\mu i}, \,\,\rm{and}\,\,V_{e i} $) mentioned in the Eq.~\ref{eq:light-heavy}.

\begin{figure*}[t!]
\begin{center}
\includegraphics[scale=0.42]{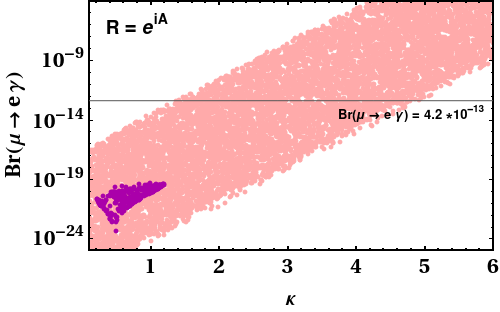}
\includegraphics[scale=0.42]{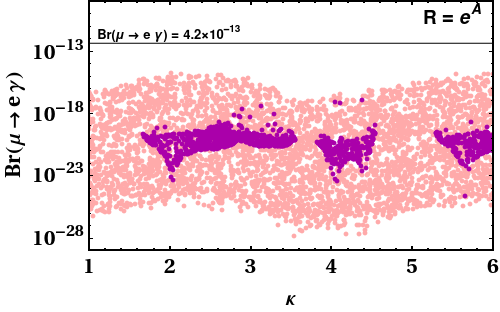}   \\
 \includegraphics[scale=0.42]{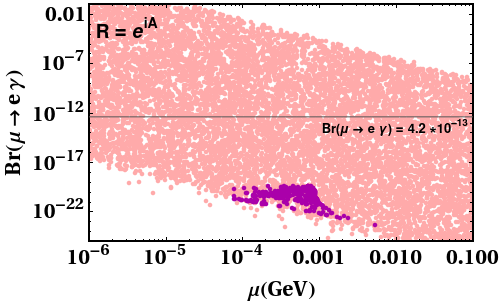}
\includegraphics[scale=0.42]{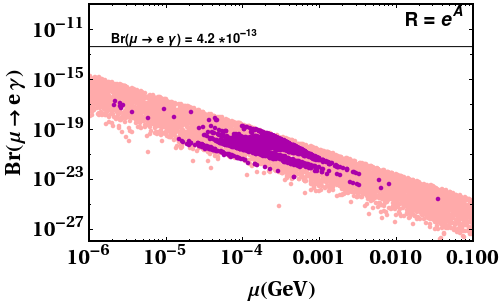}
\caption{BR($\mu \rightarrow e \gamma$) plots as a function of the $R$-matrix element for the complex (left) and real (right) cases. The grey line here indicates the present sensitivity on the said BR reported by {\bf MEG} \cite{MEG:2016leq}. The regions shown by magenta points indicate the obtained BR using the required parameter space for leptogenesis, demonstrated in the figures appearing in the later sections. However, {\bf MEGII} has a goal of increased sensitivity with an order of magnitude to the level of
$6 \times 10^{-14}$ \cite{MEGII:2018kmf}.}
\label{fig:br}
\end{center}
\end{figure*}
The discussion on the desired BR can be followed from the parameter choices and their respective ranges in the ISS model. It is clear that, the light-heavy mixing is controlled by both the LNV scale $\mu$ and $\kappa$. The numerical evaluation of the BR involves the parameter-range choices following Fig.~\ref{fig:kappa}. And it is quite imperative to understand that the mixing has a decreasing tendency (w.r.t. the traditional $\mu$ scale in ISS $\mathcal{O}(keV)$) with increasing $\mu$-scale. The exponential rise in $\kappa$ can still lead to a large branching ratio.
With the help of the range for $\kappa$ as obtained from the analysis demonstrated by the Fig.~\ref{fig:kappa}, we calculate the BR$(\mu \rightarrow e \gamma)$ and compare our result with the present bound which is BR$(\mu \rightarrow e \gamma ) \, < 4.2 \times 10^{-13}$ (Refs.~\cite{Aubert:2009ag,Adam:2013mnn,MEG:2016leq}), and the future sensitivity as BR$(\mu \rightarrow e \gamma ) < 6 \times 10^{-14}$. The following expression have been used for the calculation of BR for the aforementioned LFV decay (see \cite{Abada:2014vea,Korner:1992zk}). In the expression for BR we denote $\alpha,\,\,\text{and} \,\, i$ as the lepton flavor and light, heavy neutrino generation indices. 
\begin{equation}\label{eq:light-heavy}
\text{BR} (\mu \rightarrow e\gamma) = \frac{\alpha_w^3 s_w^2}{256 \pi^2}\frac{m_\mu^5}{M_W^4}\frac{1}{\Gamma_\mu} \Big| \sum_i^9 V^*_{\mu i} V_{e i}G(y_i) \Big|^2,
\end{equation}
where, $\alpha_w = g_w^2/ 4\pi$ and $s_w^2 = 1-(M_W/M_Z)^2$ along with the loop function $G(y)$ having the following form,
\begin{equation}
G(y) = - \frac{2 y^3 + 5 y^2 - y}{4(1-y)^3}- \frac{3y^3}{2(1-y)^4}\text{ln} y , \,\,\,\text{with}\,\, y_i =\frac{m_i^2}{M_W^2}.
\end{equation}
Here, $M_W$ and $M_Z$ imply the masses of the W and Z bosons that participate in the loop diagrams of the flavor violating decay of our interest. One denotes $\Gamma_\mu$ as the decay width of the relevant decay. In the above equation $m_i$ stands for both the active and all the sterile neutrino mass states and $V$ being the NU mixing matrix as defined earlier. We would like to refer the reader to Sec. \ref{sec:CI} for the construction of such mixing matrix\footnote{We emphasize here that Unitarity-violation scale in the ISS for the present forms of the $R$-matrix is well taken care of in the entire calculation. This can also be understood by analysing the fact that an increase in the $\mu$ scale and a rise in $\kappa$ simultaneously assist us to keep the scale of NU of the ISS intact.}. The contributions of $\kappa$ and $\mu$ to the light-heavy mixing ($V$) and later in the BR enter through the construction of $m_D$ derived simply from the term $Y_\nu \,v$ which is consisted of the parameter $\mu \,, \text{and}\, \kappa$.

It is evident from the Fig.~\ref{fig:br} that, the first choice of the $R$ matrix in the CI parametrization can assist in obtaining the order of the BR to a magnitude which is pretty close to the present \cite{MEG:2016leq} and future \cite{MEGII:2018kmf} sensitivities as we have mentioned above. However, the second choice seems to be little less competitive in this regard. For $R = e^{iA}$ case since the light-heavy mixing carries a hyperbolic kind of rise via $m_D$, the BR is enhanced by several order of magnitude with increase in $\kappa$. However, in the later case a repetitive pattern is observed for the BR when plotted as a function of $\kappa$ (in the right of Fig.~\ref{fig:br}). For the complex case one can notice the smallest value of $\kappa$ (which is around $1.2$) that is required in order to get the required and largest possible branching ratio. However for the real $R$-matrix we do not have such finding, rather having some values after certain interval which apply to the pattern obtained as shown in the figure (left of Fig.\ref{fig:br}).  A detailed reporting of these findings can be found in the table~\ref{tab:final}. One can realize that, the branching ratio is a decreasing function of the $\mu$ scale as evident from Fig.~\ref{fig:br} and understood also from Eqs.~\ref{eq:CI_1} and \ref{eq:CI_2}. 
The results of this  section gives us further direction on the range of $\kappa$ and the $\mu$ scale for the calculation of lepton and baryon asymmetry in the present theoretical background. In this figure we have also reported the amount of BR (by magenta points) that is sensitive to the $\eta_B$ constraint. This amounts to the reduction of parameter space both in the context of $\kappa$ and $\mu$. For the complex $R$-matrix the order of magnitude of the BR which is calculated using the $\eta_B$ satisfied region is more suppressed, while that is for the later choice of $R$ still larger if compared. The point we want to make is that, the combined constraint on $\mu$ and $\kappa$ which yield the observed $\eta_B$ is not sufficient to generate the desired order of magnitude for the BR. This conclusion applies to both the choices of the $R$-matrix. In the following sections we report the constraints on the ranges of $\mu$ and $\kappa$ obtained from lepton and baryon asymmetry.

\subsection{Check on the resonant condition $\Delta N/ \Gamma_N = 1$}\label{sec:analysis2}
The Pilaftsis-Underwood resonant condition \cite{Pilaftsis:2003gt} which implies the equality between the pseudo-Dirac  mass splitting and the decay rate of the lightest pseudo-Dirac state demands the condition $\Delta N/ \Gamma_N = 1$ to be true. Here, $\Delta N \,=\, (M_{N_2}^2 \,-\, M_{N_1}^2) / M_{N_1}$ and $\Gamma_{N_i}$ is the decay width of the i-th pseudo-Dirac state. This implies that, on meeting this criteria the lepton asymmetry gets resonantly enhanced, even to order $1$. In the ISS scheme the pseudo-Dirac neutral states, are splitted through the $\mu$ scale as also evident from the Eq.~\ref{eq:splitting}. It is clear from the literature \cite{Dolan:2018qpy} that for a specific $\mu$
\begin{figure*}[t!]
\begin{center}  
\includegraphics[scale=0.28]{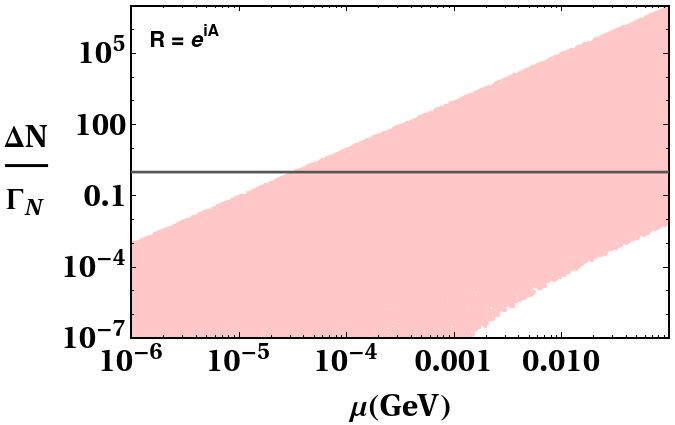}
\includegraphics[scale=0.295]{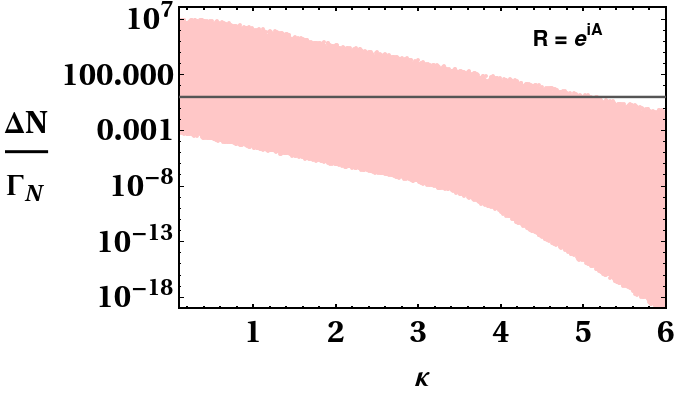}
\includegraphics[scale=0.28]{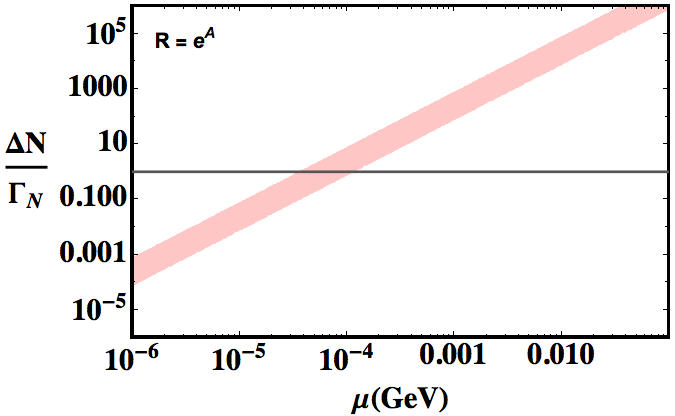} 
\includegraphics[scale=0.28]{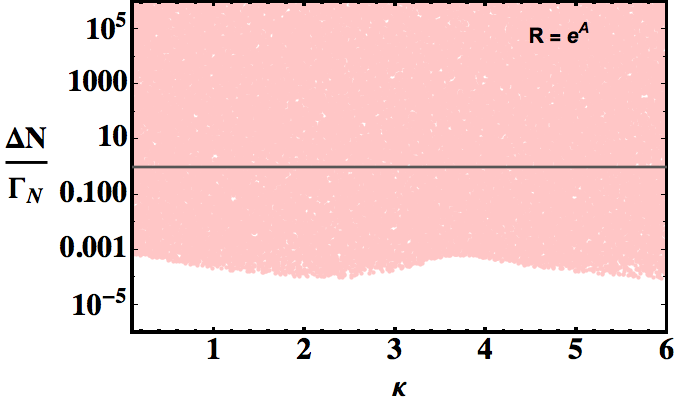} 
\caption{Shows the ISS parameter space satisfying the resonance condition, implying $\Delta N/ \Gamma_N = 1$. As evident $\Delta N/ \Gamma_N$ can reach $1$ for a wider range of $\mu$ which is assisted by the parameter $\kappa$. It is the combination of $\mu$ and $\kappa$ which simultaneously account for a wider resonance regime. However, to get a large CP asymmetry we do not require $\mu = 10^{-4}$ or even larger, as $10^{-4}< \mu < 10^{-2}$ is pretty enough to control the overall washout of the lepton asymmetry. The thickness of these bands are due to the random scan over the parameters $\kappa,\, \mu \,\text{and}\, \delta$.}
\label{fig:resonant}
\end{center}
\end{figure*}
value the lepton asymmetry attains a maximum value although, not enough to account for the observed baryon asymmetry. Authors in~\cite{Dolan:2018qpy} claimed that the conventional $\mu$ scale in ISS introduces a rise in the magnitude of the washout amount ($K$) which erases the produced lepton asymmetry which results into a smaller order of magnitude for the observed baryon asymmetry. In this work we have potentially overcome this problem in the light of these unorthodox structures of the $R$-matrix, the discussion on which is as follows. Fig.~\ref{fig:resonant} provides a realization of how these particular $R$-matrices work in meeting the resonant criteria for a wider range of the $\mu$ scale. For both the cases here, one can obtain a range of $\mu$ which not only satisfies the resonant criteria but also ensures the achievement of successful leptogenesis with the Dirac CP violation in the ISS scenario. However, with the traditional LNV scale implying $\mu = 10^{-6}$\,GeV it is almost impossible to have a successful leptogenesis even with the presence of the exponential increase in $\kappa$. Fig.~\ref{fig:kappa} gives a quantitative estimation on the allowed range of $\kappa$ for various $\mu$ scales along with the case when $\mu = 10^{-6}$GeV. Therefore the resonant condition is examined considering the pseudo-Dirac mass at $1$~TeV with $\kappa$ varying in the range $0.1 - 6$\footnote{While doing this we vary $\delta$ from $(0\, - \,2\pi)$, although the presence/absence of non-zero $\delta$ does not affect the $\Delta N/ \Gamma_N = 1$ criteria.}. For the assumed neutrino mass ordering we fix the lightest neutrino mass ($m_1$) to be $0.001$\,eV as one of the primary inputs along with the best fit central values of the oscillation parameters mentioned earlier. From Fig.~\ref{fig:resonant} one can see the resonant region in terms of the ISS parameter ($\mu$) and $\kappa$. From this figure we realize that the resonant region in terms of $\mu$ also has received a sizeable extension for the case with $R \,=\, e^{i {\bf A}}$. This is caused by the appearance of $\kappa$ along with it's hyperbolic rise appearing in the Yukawa coupling present in the decay width of the pseudo-Dirac pair. However, for $R \,=\, e^{\bf A}$ only a certain range for the $\mu$ scale ranging from ($5 \times10^{-5}\,-\, 10^{-4}$~GeV) and
the entire input range of $\kappa$ together meet the criteria $\Delta N/ \Gamma_N = 1$.                               
%
\subsection{Lepton asymmetry and washout when $R = e^{i{\bf A}}$}
\begin{figure*}[h!]
\begin{center}
\includegraphics[scale=0.31]{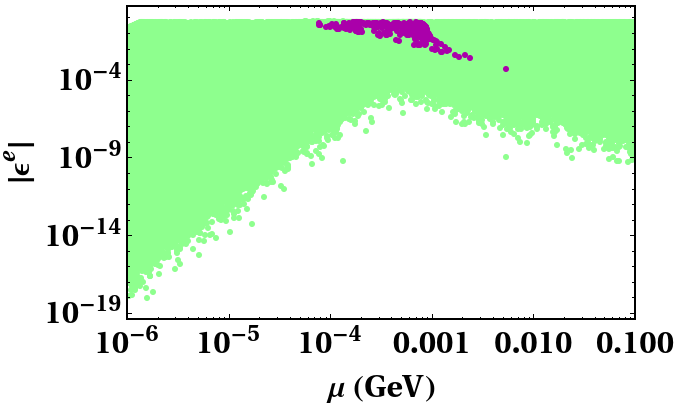}
\includegraphics[scale=0.31]{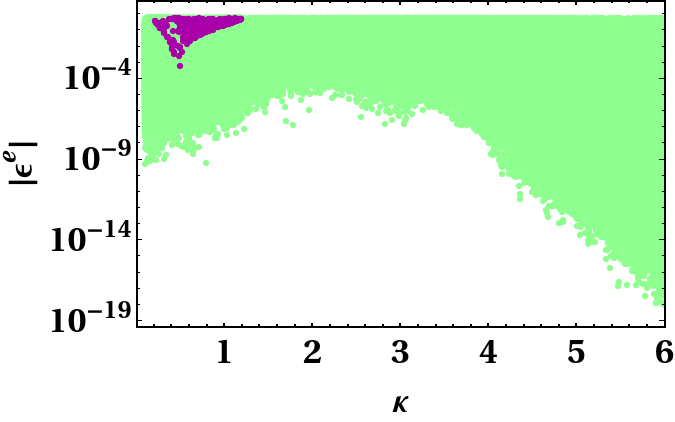}\\
\caption{The left panel shows the variation of lepton asymmetry due to electron flavor w.r.t. $\mu$, whereas the right panel presents the variation w.r.t. the parameter $\kappa$ for $R\,=\,e^{i {\bf A}}$. By green points we show the total yield of lepton asymmetry along the Y-axis, obtained for the overall random scan for the parameters presented along the X-axis. Magenta points represent the required ranges of the input  parameters which satisfy the Planck bound on $\eta_B$.}
\label{fig:cpcomplex}
\end{center}
\end{figure*}

\begin{figure*}[h]
\begin{center}
\includegraphics[scale=0.3]{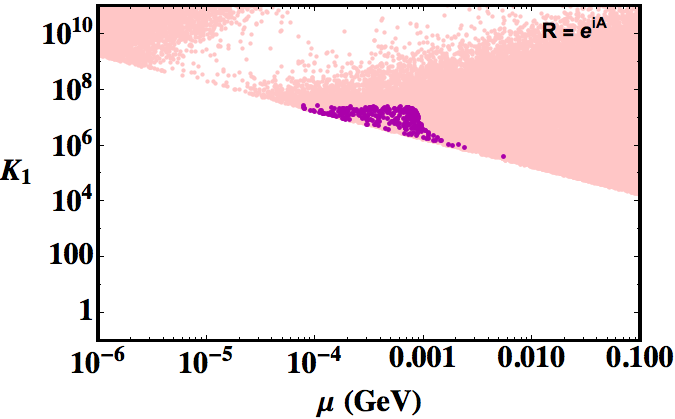}
\includegraphics[scale=0.3]{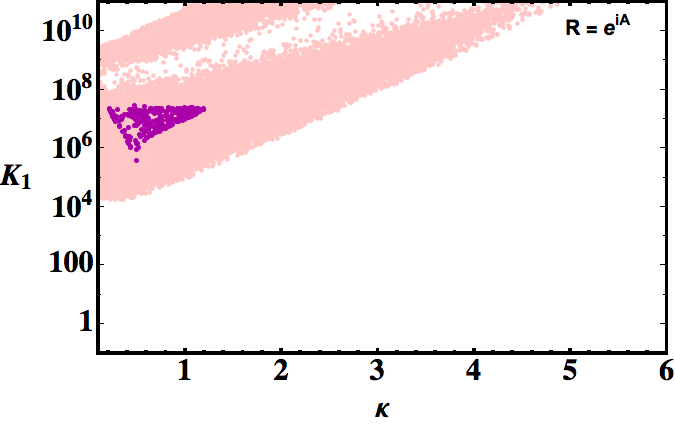}
\caption{Shows the collective influence of the LNV scale ($\mu$ on the left) and the matrix element $\kappa$ (on the right) on the total washout ($K_1$) of the asymmetry produced by the pseudo-Dirac state $N_1$. Here, the highlighted regions by magenta color indicate the amount of washout which if becomes larger will not account for the observed $\eta_B$ order. This region also determines the appropriate range for the $\mu$ and $\kappa$ parameter space one requires for leptogenesis in the ISS.}
\label{fig:washcomplex}
\end{center}
\end{figure*}

Since, the magnitude of lepton asymmetry is controlled by the parameter $\kappa$ through it's appearance in the Yukawa couplings ($Y^\nu_{ij}$) it is crucial to have a clear vision on their relation which is presented in the Fig.~\ref{fig:cpcomplex}. This figure evinces the dependency of the lepton asymmetry due to the electron flavor w.r.t. the driving parameters responsible for the
order of magnitude rise (or fall) of $Y^\nu_{ij}$\footnote{We mention here that, the lepton asymmetry associated with the other two flavors ($\mu,\,\tau$) of leptons behave analogously to $\epsilon^e$. That is the reason for keeping the discussion on lepton asymmetry relevant only for the electron flavor. This consideration will also be kept for the case study related to the later choice of the $R$ matrix.}.  The left(right) plot of this figure shows the asymmetry as a function of $\mu$($\kappa$). Here we will see how the final parameter space is decided by the asymmetry ($\epsilon_1^\ell$) and its washout such that finally these two process collectively amount to  the observed baryon to photon ratio ($\eta_B = (6.04 \,-\, 6.2) \times 10^{-10}$). On the other hand, the amount of washout, quantified by the parameter $K$ (also $\kappa_\ell^{\rm eff}$) also becomes so large that majority of the produced asymmetry gets erased, which led to a reduction in the order of magnitude of the baryon asymmetry. The discussion of how a not-very-huge washout can be obtained is provided in the explanation regarding Fig.~\ref{fig:washcomplex}. Due to the exponential rise of the Yukawa couplings the lepton asymmetry receives a huge enhancement (even $\epsilon_1^\ell >1$) which is not realistic. This happens for certain combinations of the $\kappa$ and $\mu$ parameter space, that we exclude and do not present here. The magenta region in this figure indicates the amount of lepton asymmetry required to reproduce the observed baryon asymmetry. Whereas the green region shows the net amount of asymmetry that we obtain with the present consideration of the input parameter space chosen for numerical analysis. One can notice the order of lepton asymmetry presented by the magenta points and that is ranging from $4 \times 10^{-4}\,-\,1$. This order clearly indicates the achievement of resonant criteria leading to a resonant enhancement of the asymmetry (Pilaftsis-Underwood resonance). Much smaller values of asymmetry are also obtained, but those are inadequate for the reach of the $\eta_B$ criteria. It is also to note that, even though the entire region of $\kappa$ assists in getting a resonant regime for leptogenesis, only a small portion can actually control the interplay between the generated asymmetry and the washout. From this figure we see that only smaller $\kappa$ (in a range from $0.1 - 6$) and larger $\mu$ values (from a range of $10^{-6} - 0.1$~GeV) are allowed which collectively leads to a considerable fall of the order of magnitude of the washout parameter which is otherwise impossible in a pure ISS scenario. The constraints on $\mu$ and $\kappa$ mainly come from the washout parameter, which can be understood from the following discussion related to Fig.~\ref{fig:washcomplex}. 
We explain here the underlying reason for the huge washout that is encountered in the present theoretical background. The rise of the Yukawa coupling eventually leads to an enhancement of all the parameters it is involved in. The order of the Yukawa coupling here is guided by the ISS mass scales such as, $\mu\,\,\text{and}\,\,M_N$ along with the exponential factor comprising of the parameter $\kappa$. As evident from the Eq.~\ref{eq:CI_1} an increase in $\kappa$ and a further decrease in $\mu$ scale (from $10^{-6}$\,GeV) collectively can enhance the magnitude of Yukawa coupling which in turn can engender a huge washout ($K$) along with a lepton asymmetry of order $1$\footnote{As mentioned before, even one finds the lepton asymmetry order to be more than 1, leading to an unrealistic situation. Nevertheless, we exclude those parameter space from the numerical scanning.}. Therefore, it is understandable that a larger $\mu$ scale can decrease the order of magnitude of $K$ from what it is for $\mu = 10^{-6}$~GeV. This can be realized from the following equation which is derived from $K_1 = \Gamma_1/H$. Using Eq.~\ref{eq:CI_1} in the expression for the  decay width $\Gamma_i = \frac{M_i}{8\pi}(Y_\nu Y_\nu^\dagger)_{ii}$ and incorporating it into $K_1 \,=\, \Gamma_1/H$ one can write the analytical expression for $K_1$ in terms of the mass parameters involved in the ISS mechanism and the parameter $\kappa$ as given below.
\begin{equation}\label{eq:Kcomplex}
K_1\approx \frac{m_\nu M_N}{\mu} M_{\rm Pl}  \left(0.926 \cosh (2 \sqrt{3}  \kappa)+0.073\right)
\end{equation}
The above expression for the washout has been derived using the numerical inputs for the neutrino parameters mentioned in Eq.~\ref{eq:nudata}. On the other hand, a relatively smaller order of washout can also be attained for a small value of $\kappa$ and a larger $\mu$ order. Then one has to ensure the fulfilment of a resonantly enhanced lepton asymmetry criteria which is supposed to be brought about by such combination of $\mu$ and $\kappa$ parameter values. In Fig.\,\ref{fig:washcomplex} we present the washout factor w.r.t. the driving parameters $\mu$~(left) and $\kappa$~(right). We notice here that from the beginning of $\kappa$ value (i.e. 0.1), $K_1$ goes as high as $10^{6 \,-\,9}$. The situation becomes even worse for $\kappa$ values larger than 1.5 or so. In this situation one must have $\epsilon \approx \mathcal{O} (10^{-4}\,-\,1)$ to protect the $\eta_B$ criteria that we obtain due to successful resonance as evinced from Fig.\,\ref{fig:cpcomplex} and Fig.\,\ref{fig:resonant}. However, within the $\mu$ scale of our interest the region with larger $\mu$ values only (shown by the magenta points in Fig.\,\ref{fig:washcomplex}) turns out to be essential to have an appropriate amount of washout as also discussed in the context of Fig.\,\ref{fig:cpcomplex}. 
\begin{figure*}[t]
\begin{center}
\includegraphics[scale=0.3]{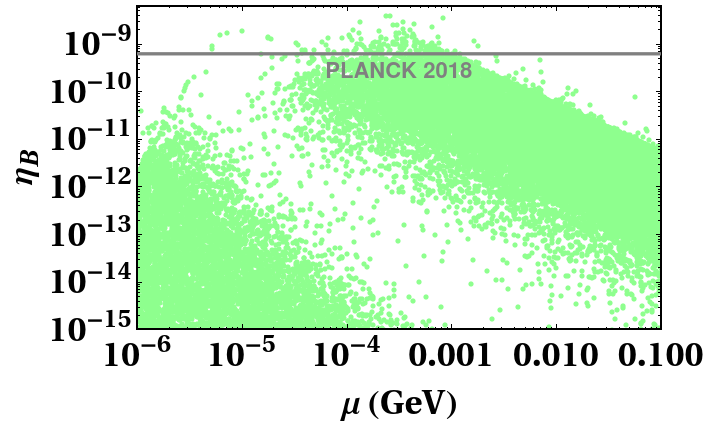}  
\includegraphics[scale=0.3]{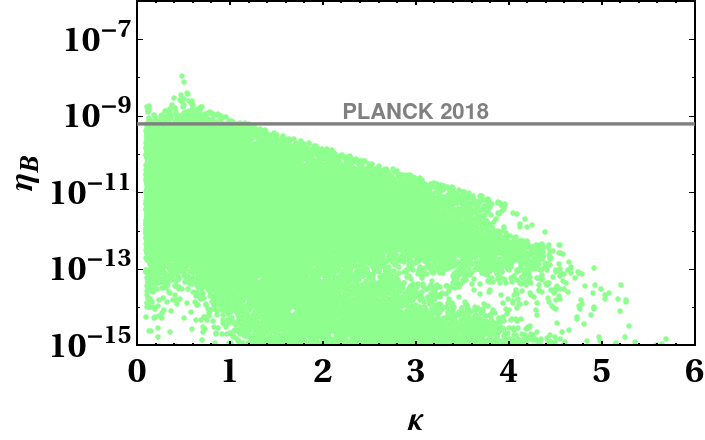} \\  
\includegraphics[scale=0.29]{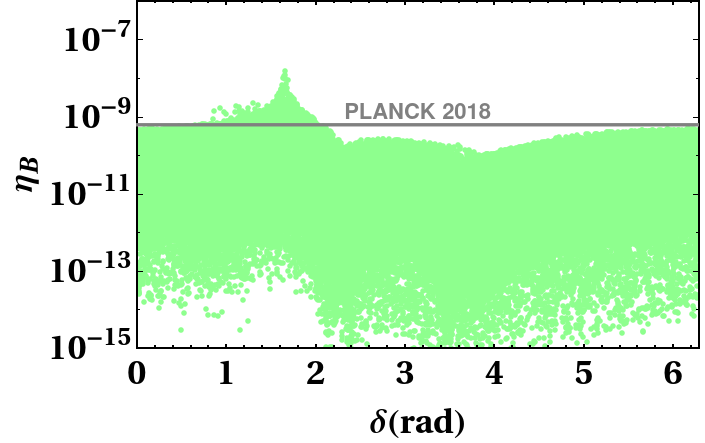}  
\caption{Represents the variation of $\eta_B$ w.r.t. the $\mu$, $\kappa$ and Dirac CP phase $\delta$ for the case of the rotational matrix $R \, =\, e^{i {\bf A}}$.}
\label{fig:etabcomplex}
\end{center}
\end{figure*}

In Fig.\,\ref{fig:etabcomplex} the final baryon to photon ratio, hence reproduced is presented w.r.t. the Dirac CP phase, the LNV scale ($\mu$) and the $R$ matrix element ($\kappa$). Here we present the total output of the baryon to photon ratio obtained out of the random scan over the aforementioned parameters, which are plotted along the Y-axes of these plots. The grey band in the figure indicates the bound on $\eta_B$ as reported by the recent PLANCK data \cite{Planck:2018vyg}. This figure also naturally indicates the same constraints on the parameters $\mu$ and $\kappa$ as depicted by the magenta points in Fig.\,\ref{fig:cpcomplex}. One significant point we want to make here is that, the input range of $\delta$ receives severe restriction (in an input range of $0 \,-\, 2\pi$) from the requirement of successful leptogenesis in this set up.

In the context of $R = e^{i{\bf A}}$ one should keep in mind the existence of an alternate source of Yukawa complexity (i.e. $Y_{\nu_{ij}}$ being complex), which is other than the Dirac CP phase. This alternate source of complex Yukawa coupling is essentially the parameter $\kappa$. We explore this possibility, in view of achieving successful leptogenesis and present the relevant results in the Appendix \ref{appendix_complex}. There also we have obtained a narrow range of ISS ($\mu$) and $\kappa$ parameter space which can yield the observed baryon to photon ratio. However, in this case the restriction on $\mu \,-\,\kappa$ ranges is much more severe.
\subsection{Lepton asymmetry and washout when $R = e^{{\bf A}}$}
\begin{figure*}[t]
\begin{center}
\includegraphics[scale=0.37]{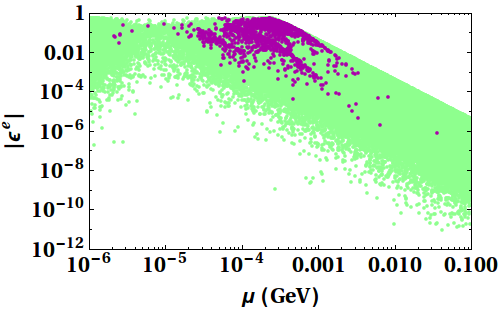}
\includegraphics[scale=0.37]{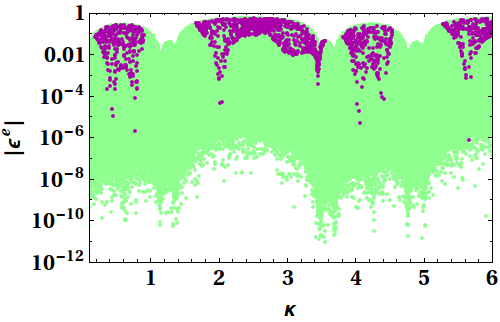} 
\caption{Lepton asymmetry due to electron flavor versus $\mu$ (left panel) and $\kappa$ (right panel) for $R\,=\,e^{\bf A}$. Green region encodes the total yield of lepton asymmetry irrespective of whether it satisfies the $\eta_B$ constraint or not. Magenta regions represent the final parameter space which accounts for the observed $\eta_B$.}
\label{fig:cpreal}
\end{center}
\end{figure*}

It is evident from Eq.~\ref{eq:CI_2} that the later choice of the rotational matrix essentially requires a low energy CP phase for leptogenesis to happen as there is no alternate source of complex Yukawa coupling in the theory. For the calculation of lepton asymmetry with this consideration of $R$ matrix, we repeat the entire procedure analogously to the previous case described in the earlier section. We have varied the parameters $\mu$, $\kappa$ and $\delta$ in their respective ranges as mentioned in the previous section and calculate the lepton and baryon asymmetry parameter. In Fig.\,\ref{fig:cpreal} we present the variation of lepton asymmetry due to the electron flavor ($\epsilon_1^\ell$) w.r.t. $\kappa$\,(left) and $\mu$\,(right), with the view that the order of lepton asymmetry is controlled by these two parameters. Analogous to the previous case with $R = e^{i {\bf A}}$ here also we see a order of magnitude rise in the Yukawa coupling which enhances $\epsilon_1$ even to order $> 1$. This leads to exclude some ranges of the input parameter space associated with $\mu$ and $\kappa$. The green regions in these plots indicate the total yield of lepton asymmetry with magnitude less than or equal to $1$ ($\epsilon \leq 1$). This green region receives further restriction when we impose the $\eta_B$ constraint on the total yield of $\epsilon_1$ which has been shown by magenta points. Unlike the previous case (with $R = e^{i{\bf A}}$) here we see that, the behaviour of lepton asymmetry with varying $\kappa$ is of periodic kind. Due to this periodicity there exists more than one allowed ranges of $\kappa$ which satisfy the $\eta_B$ criteria. However the allowed range for $\mu$ is considerably wider in this case. A comprehensive list comprizing of these findings on the allowed parameter space can be found in table \ref{tab:final}. 

\begin{figure*}[t]
\begin{center}
\includegraphics[scale=0.42]{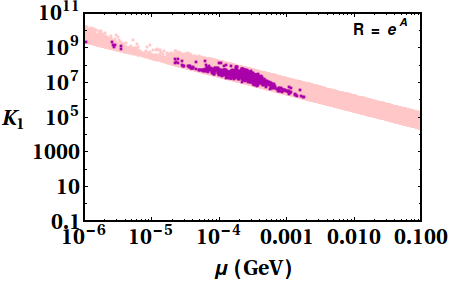}  
\includegraphics[scale=0.42]{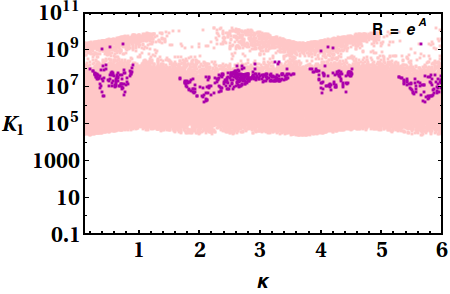}
\caption{Variation of the washout parameter w.r.t. $\mu$ and $\kappa$. The magenta regions imply the parameter space needed to see the observed baryon asymmetry despite having the huge washout.}
\label{fig:washout_K2}
\end{center}
\end{figure*}

For this choice of $R$ matrix the washout parameter behaves differently with $\kappa$ and $\mu$. The order of washout here is essentially decided by the order of the LNV scale $\mu$ and found to be very much reluctant with increase in $\kappa$ value. This can be realized from the Eq.~\ref{eq:Kreal} and Fig.~\ref{fig:washout_K2}. Following is the analytical expression presenting the overall washout as a function of $\kappa$ and other mass parameters associated with the ISS model with $R= e^{A}$. Putting the expression for $Y_\nu$ from Eq.~\ref{eq:CI_2} in $K_1 = \Gamma_1/H$ and using the neutrino data from Eq.~\ref{eq:nudata} we have, 
\begin{align} \label{eq:Kreal}
\begin{split}
K_1 \approx & \frac{m_\nu M_N}{\mu} M_{\rm Pl}\Big( 3.32+ 0.8 \sin(\sqrt{3} \kappa)(1\, - \cos (\sqrt{3} \kappa))-\cos^2(\sqrt{3}  \kappa)-1.58 \cos (\sqrt{3}  \kappa)\\ 
&\qquad+0.18 \cos (2 \sqrt{3}  \kappa)\Big)
\end{split}
\end{align}
The above expression justifies that, since the Yukawa coupling varies nearly periodically with the parameter $\kappa$, we do not see a sharp rise in the washout unlike the previous case.  Fig.\,\ref{fig:washout_K2} shows the plots for overall washout factor w.r.t. $\kappa$ and $\mu$ values. The left panel of this figure proves that like the former case the overall washout decreases with an increase in $\mu$ from the traditional ISS $\mu$ scale (which is $\mathcal {O} (10^{-6})$GeV). Whereas the washout parameter behaves almost independently with $\kappa$ through out the entire range  from (0.1 \,-\,6). It can also be understood from the Eq.~\ref{eq:Kreal}. The sinusoidal dependence of the Yukawa coupling on $\kappa$ does not bring an abrupt rise in the washout w.r.t. the suppression caused by an increasing $\mu$. Thus, the acceptable parameter space for the baryon asymmetry constraints presented by the magenta points allow the $\kappa$ value starting from $1.6$ to the possible upper bound set by the $M_N - \kappa$ plane (see Fig.\,\ref{fig:kappa}). However, in the lower panel of this figure one can see that, there is no such preference for $\mu$ which might have required to have a desired amount of lepton asymmetry and washout. The entire range of $\mu$ from $10^{-5} \,-\, 10^{-4}$\,GeV can easily accommodate an ample amount of final baryon asymmetry for this choice of rotational matrix. 

\begin{figure*}[h!]
\begin{center}
\includegraphics[scale=0.45]{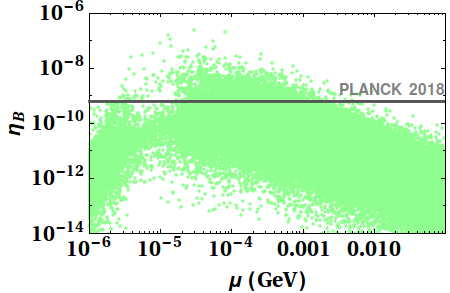}  
\includegraphics[scale=0.4]{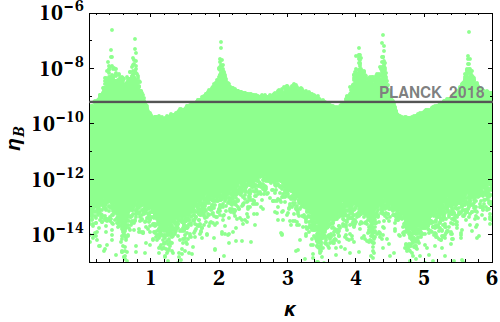}    
\includegraphics[scale=0.4]{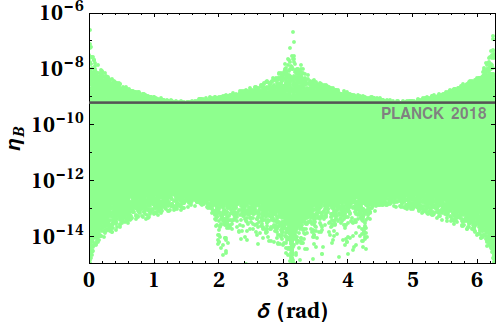}
\caption{Shows the $\eta_B$ versus $\mu$, $\kappa$ and CP phase $\delta$ for $R\,=\,e^{ A}$. The grey band indicates the recent bound on $\eta_B$ as reported by CMB observations.}
\label{fig:etabreal}
\end{center}
\end{figure*}
Constraint on $\delta$ imposed by the leptogenesis parameter space can be found from Fig.\,\ref{fig:etabreal}. The later choice of rotational matrix allow the Dirac CP phase to take some values from the range $0 \,-\, 2\pi$ after regular intervals as evinced in the first panel of this figure. It is noticed that the allowed regions are near the CP conserving values e.g., $0, \,\pi, \, 2\pi$. That is clear from The later two figures represent the dependency of the baryon asymmetry as a function of the LNV scale and the parameter $\kappa$. In this case, the imposed constraints on $\mu$ and $\kappa$ are relatively lenient and less predictive. However, it is to mention that the large values of $\kappa$ as allowed by the imposition of $\eta_B,\, \epsilon_1$ constraints at a time can be useful in order to get an enhanced BR for the process $\mu \rightarrow e\gamma$ which is found to be $10^{-18}$ for $\kappa = 2.5$ (see Fig.\,\ref{fig:br}).


\subsection{The viable parameter space for leptogenesis}
\begin{figure*}[h]
\begin{center}
\includegraphics[scale=0.31]{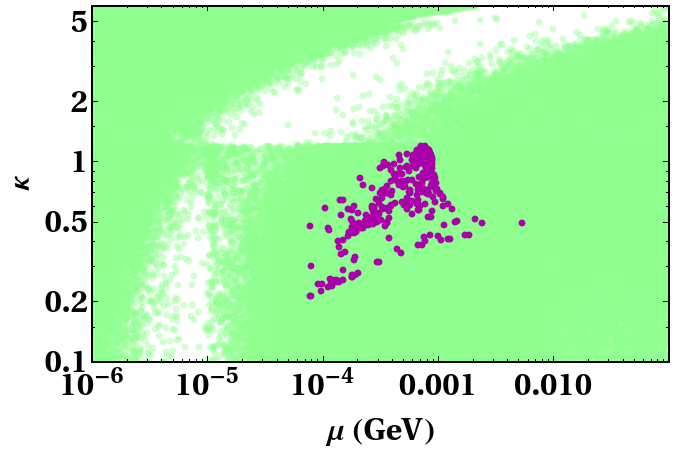}
\includegraphics[scale=0.36]{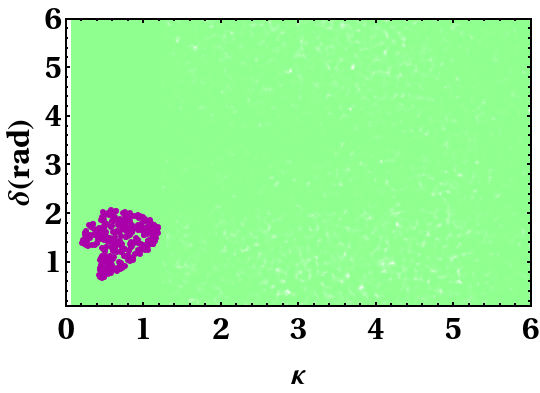}   \\ 
\includegraphics[scale=0.4]{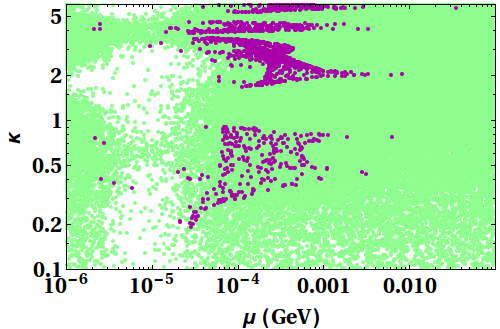}
\includegraphics[scale=0.4]{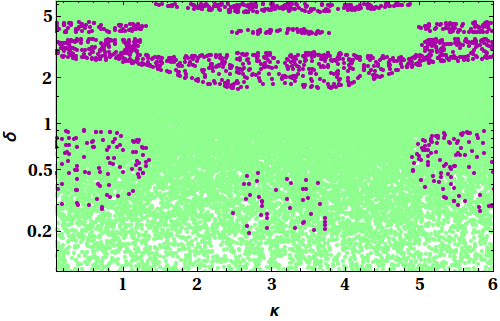} 
\caption{Constraints on the $R$-matrix parameter space and the LNV scale for the choice $R = e^{ i {\bf A}}$ (upper row) and $R = e^{\bf A}$ (bottom row). The color codes are described in the caption of figure~\ref{fig:cpcomplex}.}
\label{fig:constraints}
\end{center}
\end{figure*}

The final constraints on the parameters $\mu$ and $\kappa$ which play pivotal role in the viability of {\it Dirac phase leptogenesis} in ISS are presented in the Fig.~\ref{fig:constraints}. The restrictions on the ranges of these parameters are seen to come from three conditions, I) the perturbativity limit on the Yukawa coupling ($Y_\nu \leq |4\pi|$, Fig.~\ref{fig:kappa}), II) obtaining a close to experiment-sensitive branching ratio of the particular LFV decay (Fig.~\ref{fig:br}), and finally III) meeting the baryon asymmetry ($\eta_B$) criteria with a correct order of lepton asymmetry ($\epsilon < 1$). For clarity in understanding on the fulfilment of the above three conditions, we provide an estimation of the order of Yukawa couplings obtained for the individual $R$ matrices in table \ref{tab:coupling}. The order of magnitude for each Yukawa coupling matrices clearly indicate the fulfilment of condition I. These $Y_\nu$ matrices are evaluated using a single set of the parameters ($\mu, \,\kappa, \text{and}\,\, \delta$) presented by magenta points in Fig.~\ref{fig:constraints}. The $\delta-\kappa$ planes of the above figure (on the right) foresee different ranges for the Dirac CP phase depending on the structure of the $R$-matrix in play. With $R = e^{i {\bf A}}$ successful leptogenesis demands a maximum CP violation through $\delta$, whereas there exist more than one particular value of $\delta$ with the choice $R = e^ {\bf A}$, thus making the later scenario less predictive. The final parameter space resulting from the aforementioned constraints is provided in the table \ref{tab:final}. In this table, the first column is provided with the range of the input parameters. Then in the subsequent columns we have kept the relevant constraints related to the objective of this work, the predictions particular to each choices of $R$-matrix respectively. There we see that, the input parameter ranges get severe restrictions from several constraints mentioned in the second column. The perturbativity  condition on the Yukawa coupling gives us a range of $\mu$ and $\kappa$ which has been used for the $\eta_B$ and BR calculation. A desirable BR is achieved for a specific range of $\mu$ and $\kappa$, which in turn further gets shrunk on account of satisfying the $\eta_B$ criteria.

\begin{table}[t]
\begin{center}
\small
\begin{tabular}{| c | c | }
\hline
 Case & $Y_\nu$\\
 \hline
$ R = e^{i {\bf A} }$ & $10^{-3}
\left(\begin{array}{ccc}
 0.275\, -0.568 i & 0.474\, +0.038 i & 0.171\, +0.253 i \\
 -0.848-2.704 i & 1.202\, -1.59 i & 2.047\, +0.299 i \\
 -0.929-1.188 i & -0.106-1.269 i & 1.469\, -0.373 i \\
\end{array}
\right)$
 \\
\hline
$ R \,=\, e^{{\bf A}}$ & $10^{-3} \left(
\begin{array}{ccc}
 0.492\, -0.202 i & 0.344\, -0.143 i & 0.137\, -0.164 i \\
 1.542\, -0.057 i & 1.207\, -0.04 i & 1.14\, -0.022 i \\
 0.853\, -0.053 i & 0.476\, -0.037 i & 0.783\, -0.021 i \\
\end{array}
\right)$
 \\
\hline
\end{tabular}
\caption{An estimate for the overall Yukawa couplings which are sensitive to the leptogenesis requirement and a desired order of magnitude for the BR($\mu \rightarrow e \gamma$). This set of Yukawa matrices applies to the parameter space sensitive to the region shown by the magenta points in Fig.~\ref{fig:constraints}.}
\label{tab:coupling}
\end{center}
\end{table}

	\begin{table}[htb!]
		\renewcommand{\arraystretch}{1.4}
		\centering
		\resizebox{\linewidth}{!}{
			\begin{tabular}{|c|p{4cm}|p{4.cm}|p{4.cm}|}
				\hline 
	Input parameter ranges			 &~~~~~~~~Constraints~~~~&~~~~~~~~~~~$R=e^{i A}$~~~~&~~~~~~~~~~~~$R=e^{A}$~~~~ \\
				\hline
				$\kappa = (0.1 \,-\,6)$ & $~~~~~~~~$ $|Y_\nu| \leq \sqrt{4 \pi}$ \newline
                 $~~~~~~~$  Br($\mu \rightarrow e \gamma$) \newline
                  $~~~~~~~~~~~~~$ $\eta_B$ & $~~~~~~~~~~$0.1 - 4.5 \newline $~~~~~~~~~~$1.2 - 4.8 \newline $~~~~~~~~~~$2.24, 5.8 & $~~~~~~~~~~$0.1 - 4.78 \newline
                $~~~~~~~~~~$   0.2 - 1.4 \newline $~~~~$0.16 -0.88, 1.6 - 3.5,\newline $~~~~$3.7 - 4.5, 5.7 - 5.9 \\
				\hline
				$\mu = (10^{-6} \,-\, 10^{-1})$~GeV & $~~~~~~~~$ $|Y_\nu| \leq \sqrt{4 \pi}$\newline
               $~~~~~~~~$    Br($\mu \rightarrow e \gamma$) \newline
                 $~~~~~~~~~~~~~$  $\eta_B$ & $~~~~~~$ $10^{-6} \,-\, 10^{-1}$ \newline $~~~~~~~$ $10^{-6} \,-\, 10^{-1}$  \newline $~~~$ $8 \times 10^{-5} \,-\, 7\times 10^{-3}$ & $~~~~~$ $10^{-6} \,-\, 10^{-2}$  \newline $~~~~~~~~~~~~$ $10^{-6}$ \newline $3 \times 10^{-6} \,-\, 5\times 10^{-2}$ \\
				\hline
				$\delta = (0 \,- \,2\pi)$ radian & $~~~~~~~~$ $|Y_\nu| \leq \sqrt{4 \pi}$\newline
                 $~~~~~~~~$  Br($\mu \rightarrow e \gamma$) \newline
                 $~~~~~~~~~~~~~$  $\eta_B$ & $~~~~~~~~~~~~~~$- \newline $~~~~~~~~~~~~~~~$- \newline $~~~~~~~~~$ $0.47\,-\,2.1$ & $~~~~~~~~~~~~~~$- \newline $~~~~~~~~~~~~~~~$- \newline $0.19\,-\,0.91 , 1.76 \,-\,4.54,$\newline $~~~~~~$ $5.53\,-\,5.93$ \\
				\hline
			\end{tabular}
			}
 \caption{In the above table provided with the input parameter space used in the analysis and restrictions on them imposed by several constraints mentioned in the second column of the table.}
\label{tab:final}
	\end{table}
\section{Conclusion}\label{sec:conclusion}
In this work we have retrieved the ISS parameter space that offers {\it Dirac phase leptogenesis}. This has been accomplished by the essential involvement of two special forms of the rotational matrix present in the CI parametrisation. To guarantee the CP violation essentially sourced by the Dirac CP phase we have switched off the Majorana CP violation in the PMNS matrix. Some time ago, Ref.~\cite{Dolan:2018qpy} established that it is impossible to have a successful {\it Dirac phase leptogenesis} in a purely ISS framework, which however can be overcome when the linear seesaw is considered in addition to the ISS as a neutrino mass generation mechanism. This analysis however shows that without the linear seesaw contribution also the Dirac CP violation can be sufficient to have a successful leptogenesis in ISS owing to the presence of a special structure of the $R$ matrix. We found that the amount of lepton asymmetry that is generated can account for the observed baryon to photon ratio.

While doing so we have shown in detail the interplay of the generated lepton asymmetry and it's washout brought out by the driving parameters namely $\kappa, \,\mu, \,\, \text{and} \,\,\delta$. Having the effect of Pilaftsis-Underwood resonance, this class of $R$ matrix yields a maximum lepton asymmetry even of order $1$ for certain combinations of $\mu$ and $\kappa$ values. We have demonstrated the detailed parameter space scan associated to each individual choices of the $R$-matrix fulfilling the baryon asymmetry criteria. The final parameter spaces particular to these two cases vary substantially as explained in the previous section. For the choice with $R = e ^{ i {\bf A}}$ we could exclude a large region of the $\kappa$ and $\mu$ parameter space from simultaneously satisfying the CMB constraint on $\eta_B$ and essentially guaranteeing $\epsilon_i^\ell \leq 1$. Meeting these two constraints at a time further shrinks the final viable leptogenesis parameter space in the ISS scheme. This in turn results into a narrow region for $\kappa$ which is allowed by the three constraints mentioned in the previous section. We also have strong prediction on the $\delta$ as imposed by a successful leptogenesis and this is found to be around $(0.47-2.24)$ radian. This preference over the $\delta$ range is found to be very strict irrespective of any choice of $\mu -\kappa$ ranges. However, $\kappa$ and $\delta$ do not receive such severe restrictions for the second choice $R  = e ^{\bf A}$. For the former choice of $R$ matrix we have also performed this analysis assuming a zero Dirac CP violation in the low energy ensuring the condition $\delta,\,\alpha_1,\, \text{and}\,\alpha_2 =0$ and discussed the relevant outcome. Finally, we report that in the ISS scenario, there can exist a parameter space viable for leptogenesis which may yield a large BR for ($\mu \rightarrow e\gamma$) that is close to the vicinity of upcoming experimental sensitivities. In that case one should look after much more enhanced sensitivity for $\mu \rightarrow e\gamma$ channel with a level upto $10^{-18}$. The notable finding of this work is thus due to the unorthodox structures of the rotational matrix.  As a passing we comment that a {\it Majorana phase leptogenesis} in the ISS with the present theoretical background can foresee certain restrictions on the effective neutrino mass ($m_{\beta \beta}$) governing the $0 \nu \beta \beta$ decay, which is under investigation for an upcoming work.

\section*{Acknowledgement}
Authors are grateful to Prof. Palash Baran Pal for carefully reading this manuscript and having illuminating discussion. AM also acknowledges Prof. Gautam Bhattacharyya for his comments regarding various aspects of this work. AM also would like to acknowledge the financial support provided by SERB-DST, Govt. of India through the project EMR/2017/001434, during which period the major portion of the work was completed. 
\begin{appendices}
\section*{Appendix}\label{A1}
\section{When Dirac CP violation is also switched off for $R\,=\, e^{i {\bf A}}$}\label{appendix_complex}
\begin{figure*}[h]
\begin{center}
\includegraphics[scale=0.3]{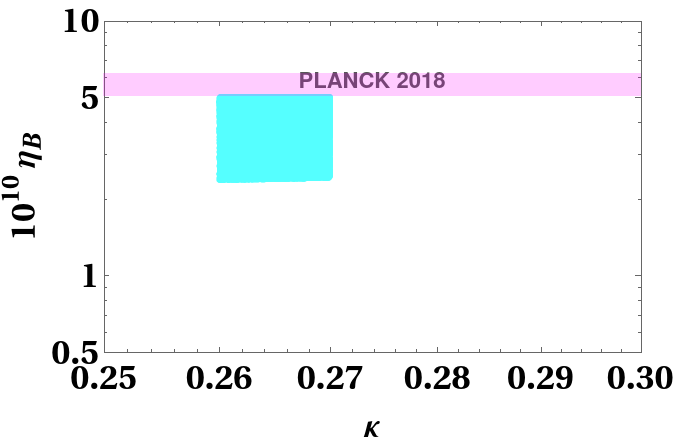}
\includegraphics[scale=0.3]{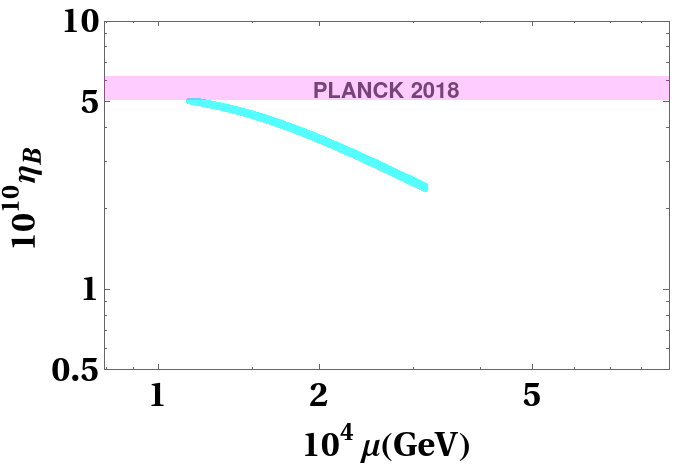}  
\caption{Shows the $\eta_B$ as a function of $\mu$ scale and $\kappa$. The cyan color indicates the total yield of asymmetry for the whole input range of parameters presented along the X-axis. The magenta patch here, is kept to show the bound on baryon to photon ratio as reported by BBN observation.}
\label{fig:deltazero1}
\end{center}
\end{figure*}

\begin{figure*}[h]
\begin{center}
\includegraphics[scale=0.3]{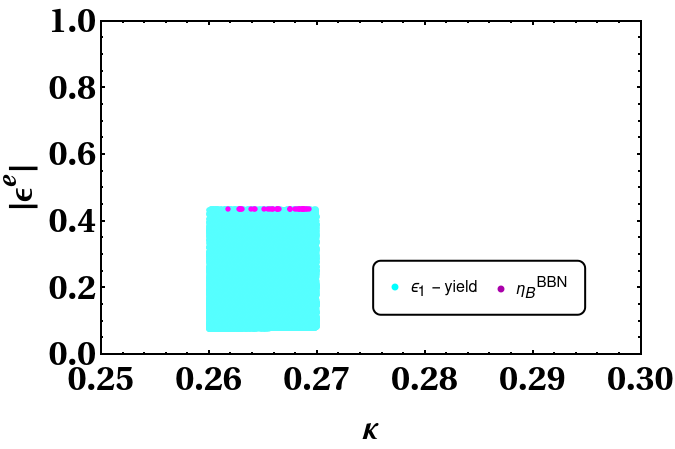}
\includegraphics[scale=0.3]{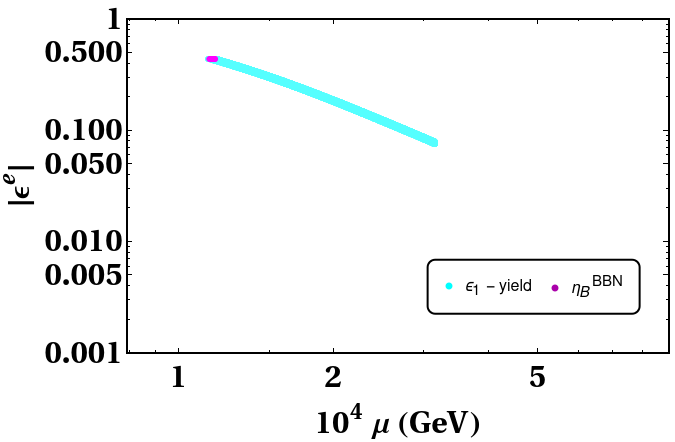}
\includegraphics[scale=0.3]{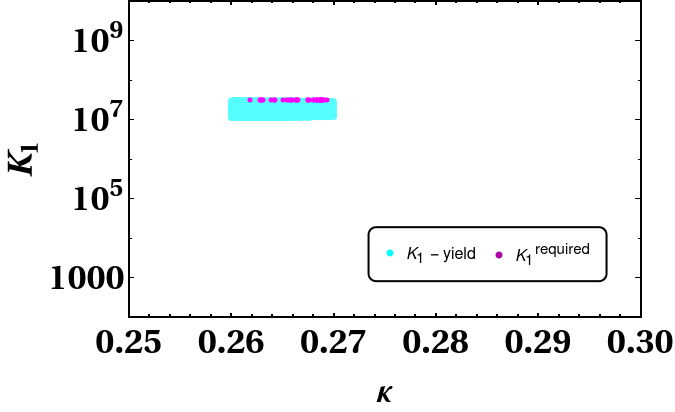}  
\includegraphics[scale=0.3]{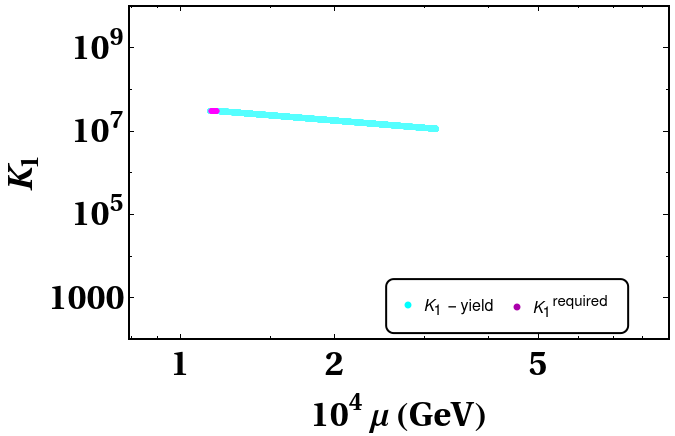}    
\caption{Variation of lepton asymmetry and its washout w.r.t. the driving parameters $\kappa$ and $\mu$. The magenta coloured region indicates the restriction on the $\kappa$ and  $ \mu$ parameter space as imposed by leptogenesis.}
\label{fig:deltazero2}
\end{center}
\end{figure*}

Here we briefly address the consequences, when at all there is no CP violation from the PMNS matrix, which is realized by having $\delta \, = \,0$. This particular analysis with $\delta \, = \,0$ is only carried out for the case when $R = e^{i A}$, as the case with the other choice of $R$ trivially predicts a zero lepton asymmetry with an exactly vanishing $\delta$. This can be understood as follows. It is evident that, due to the phase appearing in the $R$-matrix itself (Eq.~\ref{eq:CI_1}), one can assume leptogenesis to occur even after not having a low energy CP violation (implying $\delta =0, \pi, 2\pi$ etc) at all. It would be worth mentioning that, a nonzero lepton asymmetry can be obtained for a very narrow range of $\kappa$ with $\delta = 0$, which however includes  very narrow range from the bound on the baryon to photon ratio as reported by the CMB data as mentioned earlier and also in Ref.~\cite{Planck:2018vyg}. To have an insight of this fact we present here some of the significant variations of lepton and baryon asymmetry as a function of the involved parameters which play pivotal role. In Fig.~\ref{fig:deltazero1} and Fig.~\ref{fig:deltazero2}, we present the
\end{appendices}
constraints on the $\kappa -\mu$ parameter space imposed by the lepton and baryon asymmetry criteria under the assumption $\delta = 0$. For calculating the lepton asymmetry we have varied $\kappa$ in the range (0.1 - 6) along with $\mu$ from $10^{-6}\,-\, 0.1$~GeV. It is important to mention that with a vanishing $\delta$ the allowed parameter space corresponding to the involved parameters are even narrow. From an input range as mentioned above it evinces a very small part to be required for leptogenesis. Also, it is to note here that the rest of the parameter space leads to $\epsilon_i^\ell >1$ which is unrealistic. Fig.~\ref{fig:deltazero1} evinces the allowed parameter space for $\kappa$ (in the left) and the LNV scale $\mu$ (in the right) required for obtaining the amount of lepton asymmetry to be less than $1$ and which also gives rise to $\eta_B \,=\, (5.80 - 6.50) \times 10^{-10}$ coming from the BBN data \cite{Perez:2021udy}.

 Simultaneously having $\epsilon_i^\ell < 1$ and yielding $\eta_B^{\rm BBN}$ put severe restrictions on the parameter space for $\kappa$ even more in comparison to the case with $\delta \neq 0$ and finally sets an upper bound as $0.27$. This small $\kappa$ is inadequate in yielding a desired BR for $\mu \rightarrow e \gamma$ decay. On the other hand a not-very-large washout ($K_1$) which is nearly of the order of $10^8$ needs these ranges for $\kappa$ and $\mu$ as demonstrated by the lower plots in Fig.~\ref{fig:deltazero2}. Here, also we draw a similar conclusion on the large-ness of the LNV scale $\mu$ from being $10^{-6}$~GeV.  This option for leptogenesis without the Dirac CP violation is less appealing in the sense that the leptogenesis requirement for $\kappa$ is far from yielding a desired branching for $\mu \rightarrow e \gamma$ along with not satisfying the CMB bound on $\eta_B$ at all.  

\bibliography{lepto_inverse}

\providecommand{\href}[2]{#2}\begingroup\raggedright\begin{thebibliography}{10}

\bibitem{Berger:1999bg}
M.~S. Berger and B.~Brahmachari, {\it {Leptogenesis and Yukawa textures}},
  {\em Phys. Rev. D} {\bf 60} (1999) 073009,
  [\href{http://arxiv.org/abs/hep-ph/9903406}{{\tt hep-ph/9903406}}].

\bibitem{Pascoli:2006ci}
S.~Pascoli, S.~T. Petcov, and A.~Riotto, {\it {Leptogenesis and Low Energy CP
  Violation in Neutrino Physics}},  {\em Nucl. Phys. B} {\bf 774} (2007) 1--52,
  [\href{http://arxiv.org/abs/hep-ph/0611338}{{\tt hep-ph/0611338}}].

\bibitem{Joshipura:1999is}
A.~S. Joshipura and E.~A. Paschos, {\it {Constraining leptogenesis from
  laboratory experiments}},  \href{http://arxiv.org/abs/hep-ph/9906498}{{\tt
  hep-ph/9906498}}.

\bibitem{Falcone:2000ib}
D.~Falcone and F.~Tramontano, {\it {Leptogenesis and neutrino parameters}},
  {\em Phys. Rev. D} {\bf 63} (2001) 073007,
  [\href{http://arxiv.org/abs/hep-ph/0011053}{{\tt hep-ph/0011053}}].

\bibitem{Joshipura:2001ui}
A.~S. Joshipura, E.~A. Paschos, and W.~Rodejohann, {\it {A Simple connection
  between neutrino oscillation and leptogenesis}},  {\em JHEP} {\bf 08} (2001)
  029, [\href{http://arxiv.org/abs/hep-ph/0105175}{{\tt hep-ph/0105175}}].

\bibitem{Rodejohann:2002hx}
W.~Rodejohann, {\it {Leptogenesis, mass hierarchies and low-energy
  parameters}},  {\em Phys. Lett. B} {\bf 542} (2002) 100--110,
  [\href{http://arxiv.org/abs/hep-ph/0207053}{{\tt hep-ph/0207053}}].

\bibitem{Davidson:2002em}
S.~Davidson and A.~Ibarra, {\it {Leptogenesis and low-energy phases}},  {\em
  Nucl. Phys. B} {\bf 648} (2003) 345--375,
  [\href{http://arxiv.org/abs/hep-ph/0206304}{{\tt hep-ph/0206304}}].

\bibitem{Pascoli:2003uh}
S.~Pascoli, S.~T. Petcov, and W.~Rodejohann, {\it {On the connection of
  leptogenesis with low-energy CP violation and LFV charged lepton decays}},
  {\em Phys. Rev. D} {\bf 68} (2003) 093007,
  [\href{http://arxiv.org/abs/hep-ph/0302054}{{\tt hep-ph/0302054}}].

\bibitem{Frampton:2002qc}
P.~H. Frampton, S.~L. Glashow, and T.~Yanagida, {\it {Cosmological sign of
  neutrino CP violation}},  {\em Phys. Lett. B} {\bf 548} (2002) 119--121,
  [\href{http://arxiv.org/abs/hep-ph/0208157}{{\tt hep-ph/0208157}}].

\bibitem{Branco:2002kt}
G.~C. Branco, R.~Gonzalez~Felipe, F.~R. Joaquim, and M.~N. Rebelo, {\it
  {Leptogenesis, CP violation and neutrino data: What can we learn?}},  {\em
  Nucl. Phys. B} {\bf 640} (2002) 202--232,
  [\href{http://arxiv.org/abs/hep-ph/0202030}{{\tt hep-ph/0202030}}].

\bibitem{Ellis:2002xg}
J.~R. Ellis and M.~Raidal, {\it {Leptogenesis and the violation of lepton
  number and CP at low-energies}},  {\em Nucl. Phys. B} {\bf 643} (2002)
  229--246, [\href{http://arxiv.org/abs/hep-ph/0206174}{{\tt hep-ph/0206174}}].

\bibitem{Rebelo:2002wj}
M.~N. Rebelo, {\it {Leptogenesis without CP violation at low-energies}},  {\em
  Phys. Rev. D} {\bf 67} (2003) 013008,
  [\href{http://arxiv.org/abs/hep-ph/0207236}{{\tt hep-ph/0207236}}].

\bibitem{Endoh:2002wm}
T.~Endoh, S.~Kaneko, S.~K. Kang, T.~Morozumi, and M.~Tanimoto, {\it {CP
  violation in neutrino oscillation and leptogenesis}},  {\em Phys. Rev. Lett.}
  {\bf 89} (2002) 231601, [\href{http://arxiv.org/abs/hep-ph/0209020}{{\tt
  hep-ph/0209020}}].

\bibitem{Endoh:2000hc}
T.~Endoh, T.~Morozumi, T.~Onogi, and A.~Purwanto, {\it {CP violation in seesaw
  model}},  {\em Phys. Rev. D} {\bf 64} (2001) 013006,
  [\href{http://arxiv.org/abs/hep-ph/0012345}{{\tt hep-ph/0012345}}]. [Erratum:
  Phys.Rev.D 64, 059904 (2001)].

\bibitem{Molinaro:2008rg}
E.~Molinaro and S.~T. Petcov, {\it {The Interplay Between the 'Low' and 'High'
  Energy CP-Violation in Leptogenesis}},  {\em Eur. Phys. J. C} {\bf 61} (2009)
  93--109, [\href{http://arxiv.org/abs/0803.4120}{{\tt arXiv:0803.4120}}].

\bibitem{Moffat:2018smo}
K.~Moffat, S.~Pascoli, S.~T. Petcov, and J.~Turner, {\it {Leptogenesis from Low
  Energy $CP$ Violation}},  {\em JHEP} {\bf 03} (2019) 034,
  [\href{http://arxiv.org/abs/1809.08251}{{\tt arXiv:1809.08251}}].

\bibitem{Dolan:2018qpy}
M.~J. Dolan, T.~P. Dutka, and R.~R. Volkas, {\it {Dirac-Phase Thermal
  Leptogenesis in the extended Type-I Seesaw Model}},  {\em JCAP} {\bf 06}
  (2018) 012, [\href{http://arxiv.org/abs/1802.08373}{{\tt arXiv:1802.08373}}].

\bibitem{Planck:2018vyg}
{\bf Planck} Collaboration, N.~Aghanim et~al., {\it {Planck 2018 results. VI.
  Cosmological parameters}},  {\em Astron. Astrophys.} {\bf 641} (2020) A6,
  [\href{http://arxiv.org/abs/1807.06209}{{\tt arXiv:1807.06209}}]. [Erratum:
  Astron.Astrophys. 652, C4 (2021)].

\bibitem{PhysRevD.99.123508}
M.~J. Dolan, T.~P. Dutka, and R.~R. Volkas, {\it Low-scale leptogenesis with
  minimal lepton flavor violation},  {\em Phys. Rev. D} {\bf 99} (Jun, 2019)
  123508.

\bibitem{Dias:2012xp}
A.~G. Dias, C.~A. de~S.~Pires, P.~S. Rodrigues~da Silva, and A.~Sampieri, {\it
  {A Simple Realization of the Inverse Seesaw Mechanism}},  {\em Phys. Rev. D}
  {\bf 86} (2012) 035007, [\href{http://arxiv.org/abs/1206.2590}{{\tt
  arXiv:1206.2590}}].

\bibitem{Blanchet:2006be}
S.~Blanchet and P.~Di~Bari, {\it {Flavor effects on leptogenesis predictions}},
   {\em JCAP} {\bf 03} (2007) 018,
  [\href{http://arxiv.org/abs/hep-ph/0607330}{{\tt hep-ph/0607330}}].

\bibitem{Pascoli:2003rq}
S.~Pascoli, S.~T. Petcov, and C.~E. Yaguna, {\it {Quasidegenerate neutrino mass
  spectrum, mu ---\ensuremath{>} e + gamma decay and leptogenesis}},  {\em
  Phys. Lett. B} {\bf 564} (2003) 241--254,
  [\href{http://arxiv.org/abs/hep-ph/0301095}{{\tt hep-ph/0301095}}].

\bibitem{Petcov:2005yh}
S.~T. Petcov, T.~Shindou, and Y.~Takanishi, {\it {Majorana CP-violating phases,
  RG running of neutrino mixing parameters and charged lepton flavor violating
  decays}},  {\em Nucl. Phys. B} {\bf 738} (2006) 219--242,
  [\href{http://arxiv.org/abs/hep-ph/0508243}{{\tt hep-ph/0508243}}].

\bibitem{Petcov:2006pc}
S.~T. Petcov and T.~Shindou, {\it {Charged lepton decays l(i) ---\ensuremath{>}
  l(j) + gamma, leptogenesis CP-violating parameters and Majorana phases}},
  {\em Phys. Rev. D} {\bf 74} (2006) 073006,
  [\href{http://arxiv.org/abs/hep-ph/0605151}{{\tt hep-ph/0605151}}].

\bibitem{Konar:2020vuu}
P.~Konar, A.~Mukherjee, A.~K. Saha, and S.~Show, {\it {A dark clue to seesaw
  and leptogenesis in a pseudo-Dirac singlet doublet scenario with
  (non)standard cosmology}},  \href{http://arxiv.org/abs/2007.15608}{{\tt
  arXiv:2007.15608}}.

\bibitem{Mukherjee:2021hed}
A.~Mukherjee and N.~Narendra, {\it {Realizing flavored leptogenesis: a
  reappraisal through special kinds of orthogonal matrices}},
  \href{http://arxiv.org/abs/2105.14593}{{\tt arXiv:2105.14593}}.

\bibitem{Casas:2001sr}
J.~A. Casas and A.~Ibarra, {\it {Oscillating neutrinos and $\mu \to e,
  \gamma$}},  {\em Nucl. Phys. B} {\bf 618} (2001) 171--204,
  [\href{http://arxiv.org/abs/hep-ph/0103065}{{\tt hep-ph/0103065}}].

\bibitem{Arun:2021yhm}
M.~T. Arun, T.~Mandal, S.~Mitra, A.~Mukherjee, L.~Priya, and A.~Sampath, {\it
  {Testing left-right symmetry with inverse seesaw at the LHC}},
  \href{http://arxiv.org/abs/2109.09585}{{\tt arXiv:2109.09585}}.

\bibitem{PhysRevLett.56.561}
R.~N. Mohapatra, {\it Mechanism for understanding small neutrino mass in
  superstring theories},  {\em Phys. Rev. Lett.} {\bf 56} (Feb, 1986) 561--563.

\bibitem{PhysRevD.34.1642}
R.~N. Mohapatra and J.~W.~F. Valle, {\it Neutrino mass and baryon-number
  nonconservation in superstring models},  {\em Phys. Rev. D} {\bf 34} (Sep,
  1986) 1642--1645.

\bibitem{Esteban:2018azc}
I.~Esteban, M.~Gonzalez-Garcia, A.~Hernandez-Cabezudo, M.~Maltoni, and
  T.~Schwetz, {\it {Global analysis of three-flavour neutrino oscillations:
  synergies and tensions in the determination of $\theta_{23}$, $\delta_{CP}$,
  and the mass ordering}},  {\em JHEP} {\bf 01} (2019) 106,
  [\href{http://arxiv.org/abs/1811.05487}{{\tt arXiv:1811.05487}}].

\bibitem{Abe:2019vii}
{\bf T2K} Collaboration, K.~Abe et~al., {\it {Constraint on the
  matter\textendash{}antimatter symmetry-violating phase in neutrino
  oscillations}},  {\em Nature} {\bf 580} (2020), no.~7803 339--344,
  [\href{http://arxiv.org/abs/1910.03887}{{\tt arXiv:1910.03887}}]. [Erratum:
  Nature 583, E16 (2020)].

\bibitem{Acero:2019ksn}
{\bf NOvA} Collaboration, M.~A. Acero et~al., {\it {First Measurement of
  Neutrino Oscillation Parameters using Neutrinos and Antineutrinos by NOvA}},
  {\em Phys. Rev. Lett.} {\bf 123} (2019), no.~15 151803,
  [\href{http://arxiv.org/abs/1906.04907}{{\tt arXiv:1906.04907}}].

\bibitem{Rodejohann:2010zz}
W.~Rodejohann, {\it {Non-Unitary PMNS Matrix, Leptogenesis and Low Energy CP
  Violation}},  {\em AIP Conf. Proc.} {\bf 1222} (2010), no.~1 93--97.

\bibitem{Blennow:2016jkn}
M.~Blennow, P.~Coloma, E.~Fernandez-Martinez, J.~Hernandez-Garcia, and
  J.~Lopez-Pavon, {\it {Non-Unitarity, sterile neutrinos, and Non-Standard
  neutrino Interactions}},  {\em JHEP} {\bf 04} (2017) 153,
  [\href{http://arxiv.org/abs/1609.08637}{{\tt arXiv:1609.08637}}].

\bibitem{Pilaftsis:2003gt}
A.~Pilaftsis and T.~E. Underwood, {\it {Resonant leptogenesis}},  {\em Nucl.
  Phys. B} {\bf 692} (2004) 303--345,
  [\href{http://arxiv.org/abs/hep-ph/0309342}{{\tt hep-ph/0309342}}].

\bibitem{Pilaftsis:1997jf}
A.~Pilaftsis, {\it {CP violation and baryogenesis due to heavy Majorana
  neutrinos}},  {\em Phys. Rev. D} {\bf 56} (1997) 5431--5451,
  [\href{http://arxiv.org/abs/hep-ph/9707235}{{\tt hep-ph/9707235}}].

\bibitem{Flanz:1996fb}
M.~Flanz, E.~A. Paschos, U.~Sarkar, and J.~Weiss, {\it {Baryogenesis through
  mixing of heavy Majorana neutrinos}},  {\em Phys. Lett. B} {\bf 389} (1996)
  693--699, [\href{http://arxiv.org/abs/hep-ph/9607310}{{\tt hep-ph/9607310}}].

\bibitem{Xing:2006ms}
Z.-z. Xing and S.~Zhou, {\it {Tri-bimaximal Neutrino Mixing and
  Flavor-dependent Resonant Leptogenesis}},  {\em Phys. Lett. B} {\bf 653}
  (2007) 278--287, [\href{http://arxiv.org/abs/hep-ph/0607302}{{\tt
  hep-ph/0607302}}].

\bibitem{Abada:2006ea}
A.~Abada, S.~Davidson, A.~Ibarra, F.~X. Josse-Michaux, M.~Losada, and
  A.~Riotto, {\it {Flavour Matters in Leptogenesis}},  {\em JHEP} {\bf 09}
  (2006) 010, [\href{http://arxiv.org/abs/hep-ph/0605281}{{\tt
  hep-ph/0605281}}].

\bibitem{Nardi:2006fx}
E.~Nardi, Y.~Nir, E.~Roulet, and J.~Racker, {\it {The Importance of flavor in
  leptogenesis}},  {\em JHEP} {\bf 01} (2006) 164,
  [\href{http://arxiv.org/abs/hep-ph/0601084}{{\tt hep-ph/0601084}}].

\bibitem{Deppisch:2010fr}
F.~F. Deppisch and A.~Pilaftsis, {\it {Lepton Flavour Violation and theta(13)
  in Minimal Resonant Leptogenesis}},  {\em Phys. Rev. D} {\bf 83} (2011)
  076007, [\href{http://arxiv.org/abs/1012.1834}{{\tt arXiv:1012.1834}}].

\bibitem{Bambhaniya:2016rbb}
G.~Bambhaniya, P.~S. Bhupal~Dev, S.~Goswami, S.~Khan, and W.~Rodejohann, {\it
  {Naturalness, Vacuum Stability and Leptogenesis in the Minimal Seesaw
  Model}},  {\em Phys. Rev. D} {\bf 95} (2017), no.~9 095016,
  [\href{http://arxiv.org/abs/1611.03827}{{\tt arXiv:1611.03827}}].

\bibitem{Dev:2017trv}
P.~S.~B. Dev, P.~Di~Bari, B.~Garbrecht, S.~Lavignac, P.~Millington, and
  D.~Teresi, {\it {Flavor effects in leptogenesis}},  {\em Int. J. Mod. Phys.
  A} {\bf 33} (2018) 1842001, [\href{http://arxiv.org/abs/1711.02861}{{\tt
  arXiv:1711.02861}}].

\bibitem{Blanchet:2010kw}
S.~Blanchet, P.~S.~B. Dev, and R.~N. Mohapatra, {\it {Leptogenesis with TeV
  Scale Inverse Seesaw in SO(10)}},  {\em Phys. Rev. D} {\bf 82} (2010) 115025,
  [\href{http://arxiv.org/abs/1010.1471}{{\tt arXiv:1010.1471}}].

\bibitem{Kartavtsev:2015vto}
A.~Kartavtsev, P.~Millington, and H.~Vogel, {\it {Lepton asymmetry from mixing
  and oscillations}},  {\em JHEP} {\bf 06} (2016) 066,
  [\href{http://arxiv.org/abs/1601.03086}{{\tt arXiv:1601.03086}}].

\bibitem{Dev:2015wpa}
P.~S.~B. Dev, P.~Millington, A.~Pilaftsis, and D.~Teresi, {\it {Corrigendum to
  ''Flavour Covariant Transport Equations: an Application to Resonant
  Leptogenesis''}},  {\em Nucl. Phys. B} {\bf 897} (2015) 749--756,
  [\href{http://arxiv.org/abs/1504.07640}{{\tt arXiv:1504.07640}}].

\bibitem{PhysRevD.49.6394}
J.~M. Cline, K.~Kainulainen, and K.~A. Olive, {\it Protecting the primordial
  baryon asymmetry from erasure by sphalerons},  {\em Phys. Rev. D} {\bf 49}
  (Jun, 1994) 6394--6409.

\bibitem{DOnofrio:2012phz}
M.~D'Onofrio, K.~Rummukainen, and A.~Tranberg, {\it {The Sphaleron Rate through
  the Electroweak Cross-over}},  {\em JHEP} {\bf 08} (2012) 123,
  [\href{http://arxiv.org/abs/1207.0685}{{\tt arXiv:1207.0685}}].

\bibitem{Cheng:1980tp}
T.~P. Cheng and L.-F. Li, {\it {$\mu \to e \gamma$ in Theories With Dirac and
  Majorana Neutrino Mass Terms}},  {\em Phys. Rev. Lett.} {\bf 45} (1980) 1908.

\bibitem{Aubert:2009ag}
{\bf BaBar} Collaboration, B.~Aubert et~al., {\it {Searches for Lepton Flavor
  Violation in the Decays tau+- ---\ensuremath{>} e+- gamma and tau+-
  ---\ensuremath{>} mu+- gamma}},  {\em Phys. Rev. Lett.} {\bf 104} (2010)
  021802, [\href{http://arxiv.org/abs/0908.2381}{{\tt arXiv:0908.2381}}].

\bibitem{MEG:2016leq}
{\bf MEG} Collaboration, A.~M. Baldini et~al., {\it {Search for the lepton
  flavour violating decay $\mu ^+ \rightarrow \mathrm {e}^+ \gamma $ with the
  full dataset of the MEG experiment}},  {\em Eur. Phys. J. C} {\bf 76} (2016),
  no.~8 434, [\href{http://arxiv.org/abs/1605.05081}{{\tt arXiv:1605.05081}}].

\bibitem{MEGII:2018kmf}
{\bf MEG II} Collaboration, A.~M. Baldini et~al., {\it {The design of the MEG
  II experiment}},  {\em Eur. Phys. J. C} {\bf 78} (2018), no.~5 380,
  [\href{http://arxiv.org/abs/1801.04688}{{\tt arXiv:1801.04688}}].

\bibitem{Adam:2013mnn}
{\bf MEG} Collaboration, J.~Adam et~al., {\it {New constraint on the existence
  of the $\mu^+ \to e^+\gamma$ decay}},  {\em Phys. Rev. Lett.} {\bf 110}
  (2013) 201801, [\href{http://arxiv.org/abs/1303.0754}{{\tt
  arXiv:1303.0754}}].

\bibitem{Abada:2014vea}
A.~Abada and M.~Lucente, {\it {Looking for the minimal inverse seesaw
  realisation}},  {\em Nucl. Phys. B} {\bf 885} (2014) 651--678,
  [\href{http://arxiv.org/abs/1401.1507}{{\tt arXiv:1401.1507}}].

\bibitem{Korner:1992zk}
J.~G. Korner, A.~Pilaftsis, and K.~Schilcher, {\it {Leptonic CP asymmetries in
  flavor changing H0 decays}},  {\em Phys. Rev. D} {\bf 47} (1993) 1080--1086,
  [\href{http://arxiv.org/abs/hep-ph/9301289}{{\tt hep-ph/9301289}}].

\bibitem{Perez:2021udy}
P.~F. Perez, C.~Murgui, and A.~D. Plascencia, {\it {Baryogenesis via
  leptogenesis: Spontaneous B and L violation}},  {\em Phys. Rev. D} {\bf 104}
  (2021), no.~5 055007, [\href{http://arxiv.org/abs/2103.13397}{{\tt
  arXiv:2103.13397}}].

\end{thebibliography}\endgroup
\bibliographystyle{JHEP}
\end{document}